\documentclass[trackchanges,twocolumn]{aastex701}
\usepackage[T1]{fontenc}
\usepackage{times}
\usepackage{booktabs}
\usepackage{amsmath} 
\usepackage{mathrsfs} 
\usepackage{ulem}

\usepackage{soul}
\usepackage[section]{placeins}
\usepackage{enumitem}
\usepackage{multirow}

\newcommand{\OIII}{\hbox{[O\,{\sc iii}]}}

\newcommand{\Heii}{\hbox{He\,{\sc ii}}}
\newcommand{\Hei}{\hbox{He\,{\sc i}}}

\newcommand{\nH}{n_{\mathrm{H}}}
\newcommand{\PhiH}{\Phi_{\mathrm{H}}}
\newcommand{\PhiHiso}{\Phi_{\mathrm{H}}^{\mathrm{iso}}}
\newcommand{\PhiHion}{\Phi_{\mathrm{H}}^{\mathrm{ion}}}
\newcommand{\DeltaPhiH}{\Delta\log\Phi_{\mathrm{H}}}

\def\RFe{{\cal R}_{\rm Fe}}

\def\mathdotM{\dot{\mathscr{M}}}

\def\Mdot{\dot{\mathscr{M}}}

\def\M{{M}_{\bullet}}

\def\authoremaila{\href{mailto:dupu@ihep.ac.cn}{dupu@ihep.ac.cn}}

\shortauthors{F.-N. Fang et al. }

\begin{document}

\title{Anisotropic Radiation in Super-Eddington Active Galactic Nuclei: The Responsivity of Broad-Line Regions to Continuum Variability}

\author[0009-0009-9945-673X]{Feng-Na Fang}
\affiliation{State Key Laboratory of Particle Astrophysics, Institute of High Energy Physics, Chinese Academy of Sciences, 19B Yuquan Road, Beijing 100049, China; \authoremaila}
\email{fangfengna@ihep.ac.cn}

\author[0000-0002-5830-3544]{Pu Du}
\affiliation{State Key Laboratory of Particle Astrophysics, Institute of High Energy Physics, Chinese Academy of Sciences, 19B Yuquan Road, Beijing 100049, China; \authoremaila}
\email{dupu@ihep.ac.cn}

\author[0000-0001-9449-9268]{Jian-Min Wang}
\affiliation{State Key Laboratory of Particle Astrophysics, Institute of High Energy Physics, Chinese Academy of Sciences, 19B Yuquan Road, Beijing 100049, China; \authoremaila}
\affiliation{National Astronomical Observatories, Chinese Academy of Sciences, 20A Datun Road, Chaoyang District, Beijing 100101, China}
\affiliation{School of Astronomy and Space Science, University of Chinese Academy of Sciences, 19A Yuquan Road, Beijing 100049, China}
\email{wangjm@ihep.ac.cn}

\begin{abstract}

Anisotropic ionizing radiation is a natural theoretical expectation for super-Eddington accretion in active galactic nuclei (AGNs), implying that the ionizing flux incident on the broad-line region (BLR) clouds differs from that measured along the observer’s line of sight. Reverberation mapping studies have revealed systematically shortened time lags of the broad H$\beta$ emission line relative to continuum variations in super-Eddington AGNs, a phenomenon that can be naturally interpreted as a consequence of anisotropic ionizing continuum radiation. To further investigate this anisotropy, we perform a systematic analysis of the H$\beta$ line responsivity, defined as the ratio of the variability amplitude of H$\beta$ to that of the driving continuum, for both sub- and super-Eddington AGNs. By combining the effective responsivities measured relative to the driving continuum, after correcting for the continuum--emission-line delay, with photoionization calculations, we infer the ionizing radiation field received by their BLRs. We find that, in super-Eddington AGNs, the ionizing flux inferred from the H$\beta$ responsivity differs significantly from that observed along the line of sight, whereas the two are in closer agreement in sub-Eddington AGNs. These results provide compelling evidence for strong anisotropy of the incident ionizing radiation in super-Eddington AGNs. We conclude by discussing the broader implications of this anisotropy, including its potential relevance to the ``little red dots'' recently discovered by the James Webb Space Telescope.

\end{abstract}


\section{Introduction} \label{sec:intro}

Super-Eddington active galactic nuclei (AGNs)---a subclass of AGNs characterized by extremely high accretion rates---play a crucial role in the growth of supermassive black holes (SMBHs), particularly in the early Universe. Recent observations with the \textit{James Webb Space Telescope} (JWST) have revealed that a substantial fraction of the most distant quasars are likely accreting at super-Eddington rates \citep[e.g.,][]{harikene2023, yang2023, maiolino2024, maiolino2024AAp, suh2025}, further underscoring the importance of this remarkable subclass of AGNs. Nevertheless, the detailed physics of super-Eddington accretion and the corresponding structure of such AGNs remain poorly understood. 

A key theoretical prediction of super-Eddington accretion is that the accretion flow becomes geometrically thick. This arises both from the classical slim disk model \citep[e.g.,][]{abramowicz1988, wang1999, mineshige2000} and from recent numerical simulations \citep[e.g.,][]{ohsuga2011, jiang2019}. In this regime, the ionizing radiation emitted from the innermost regions of the disk is expected to be highly anisotropic \citep[e.g.,][]{madau1988, Wang2014_self, luo2015}. Consequently, a significant portion of the broad-line region (BLR) gas---especially that located near the disk mid-plane---may be shielded from the ionizing radiation \citep{Wang2014_self}. This so-called \emph{self-shadowing effect} can lead to a reduction in the sizes of BLRs \citep{Wang2014_self}. More generally, \citet{Korista1997a} pointed out that the ionizing continuum seen by BLR clouds may differ from that observed directly by the observer, highlighting the importance of considering the radiation field incident on the BLR.

Observationally, reverberation mapping (RM) is a powerful technique for determining the characteristic sizes of BLRs ($R_{\rm H\beta} = c\,\tau_{\rm H\beta}$) from the time delay ($\tau_{\rm H\beta}$) between variations in the continuum and the corresponding response of the emission lines (particularly the H$\beta$ line), where $c$ is the speed of light \citep[e.g.,][]{blandford_reverberation_1982, Peterson1993}. This method has been successfully applied to more than a hundred AGNs over the past three decades \citep[e.g.,][]{peterson_rm_1998, kaspi_reverberation_2000, bentz_lick_2009, grier_reverberation_2012, du_supermassive_2014, shen2015, barth2015, brotherton_monitoring_2020, malik2023, woo2024}. Consistent with theoretical expectations, super-Eddington AGNs have been found to exhibit BLR sizes smaller by factors of a few up to an order of magnitude compared with AGNs accreting at sub-Eddington rates, as revealed by the dedicated ``Super-Eddington Accreting Massive Black Holes'' (SEAMBH) RM campaign \citep[e.g.,][]{du_supermassive_2015, du_supermassive_2016, Du_Supermassive_2018}.

Moreover, interferometric observations with the GRAVITY instrument on the \textit{Very Large Telescope Interferometer} (VLTI) have spatially resolved the geometry of BLRs with angular resolutions of tens of microarcseconds, and have likewise reported smaller BLR sizes in super-Eddington AGNs across a wide redshift range \citep[e.g.,][]{gravity2024, abuter2024, gravity2025}. Together, both RM and interferometric measurements consistently support the presence of self-shadowing and anisotropic ionizing radiation in super-Eddington AGNs, as inferred from the systematically smaller BLR sizes.

To further investigate the anisotropic radiation in super-Eddington AGNs, it is desirable to find an indicator that is as independent of the time lag as possible and more directly reflects the strength of the ionizing radiation. In addition to the time lag, the \emph{responsivity}, defined as the fractional response of the emission-line flux to variations in the continuum, can serve as such an indicator. It is expressed as
\begin{equation}
    \eta = \frac{\Delta \log F_{\mathrm{line}}}{\Delta \log F_{\mathrm{cont}}},
    \label{equ:eta define}
\end{equation}
where $\Delta \log F_{\mathrm{line}}$ and $\Delta \log F_{\mathrm{cont}}$ denote the variabilities of the emission-line and continuum fluxes in logarithmic space \citep[e.g.,][]{Krolik1991, Pogge1992, goad1993, Korista_Goad_2004, Goad2004, cackett2006, Goad2014, Goad2015, guo2020, Du2023_cloudy}. This definition follows from the empirical relation observed between the line and continuum fluxes, $F_{\mathrm{line}} \propto F_{\mathrm{cont}}^{\eta}$ (see references above). From the perspective of photoionization calculations \citep[e.g.,][]{Korista_Goad_2004, Du2023_cloudy}, $\eta$ depends on both the ionizing flux density and the density of BLR gas clouds (actually more sensitive to the incident ionizing radiation, see the following sections). For optical recombination lines such as H$\beta$, the responsivity decreases with increasing incident ionizing flux primarily because the increasing populations of excited hydrogen levels lead to larger line optical depths and, eventually, line thermalization. Therefore $\eta$ can be used as a probe of the anisotropic radiation expected from slim disk in super-Eddington AGNs.

AGNs with RM observations offer an ideal sample for investigating the responsivity of emission lines, owing to their dense sampling cadence and high calibration accuracy. In this work, we compile the RM sample and analyze their H$\beta$ emission-line variability to assess the degree of anisotropic ionizing radiation from the central accretion disks. The paper is organized as follows. Section~\ref{sec:Measurement} describes the sample and the procedures employed to measure responsivity. Section~\ref{sec:analysis} presents the analysis and the main results. Some discussions are provided in Section~\ref{sec:discussion}, and Section~\ref{sec:Conclusion} summarizes our conclusions. Throughout this paper, we adopt a standard $\Lambda$CDM cosmology with $H_0 = 67~\mathrm{km~s^{-1}~Mpc^{-1}}$, $\Omega_{\Lambda} = 0.68$, and $\Omega_m = 0.32$ \citep{Planck2014, Planck2018}.

\section{Data and Measurement} \label{sec:Measurement}
\subsection{Sample}\label{sec:Sample}

To investigate the anisotropic radiation through the responsivity of BLRs, we compiled the most recent RM sample for H$\beta$ emission lines. The sample size of H$\beta$ RM measurements is the largest available and the quality of the data is the best to date. Our sample builds upon the collection by \cite{Du2019}, which compiles most sub-Eddington AGNs observed before 2013 \citep[e.g.,][]{kaspi_reverberation_2000, denney_diverse_2009, grier_reverberation_2012, bentz_low-luminosity_2013}, as well as super-Eddington AGNs from the SEAMBH project \citep[e.g.,][]{du_supermassive_2014, du_supermassive_2015, du_supermassive_2016, Du_Supermassive_2018} covering the period from 2012 to 2017. We supplemented this initial dataset with: (1) samples from the Monitoring AGNs with H$\beta$ Asymmetry (MAHA) project \citep{du_monitoring_2018, brotherton_monitoring_2020, bao_monitoring_2022, Zastrocky2024, Fang2025}, (2) newly published objects from SEAMBH campaign after 2017 \citep{Hu2020_PG2130, hu_supermassive_2021, Lishasha2021, Yao2024_PG2130}, (3) objects from the Sloan Digital Sky Survey RM project (SDSS-RM, e.g., \citealt[][]{shen_sloan_2019, shen2024}), and (4) data from the Seoul National University AGN Monitoring project (SAMP, \citealt{Rakshit2019}). For the SDSS-RM sample, we selected only the ``Gold'' objects with lag grades of 4 or 5 as reported in \cite{shen2024}, which exhibit the best data quality and the most accurate lag measurements (for details, see \citealt{shen2024}) {\footnote{Note that RM~645 and RM~772 are excluded from the current sample because the responsivity of their H$\beta$ emission to the continuum is significantly larger than that of other objects, likely due to distinct or unknown peculiarities in their BLR physics. Additionally, RM~118 was excluded because its corresponding continuum points prior to the H$\beta$ measurements, associated with the observed time delay, are predominantly located within large season gaps.}. Ultimately, the initial sample in this study comprises 151 AGNs with a total of 216 light curves. }

\begin{figure*}[!t]
    \centering
    \includegraphics[width=\textwidth]{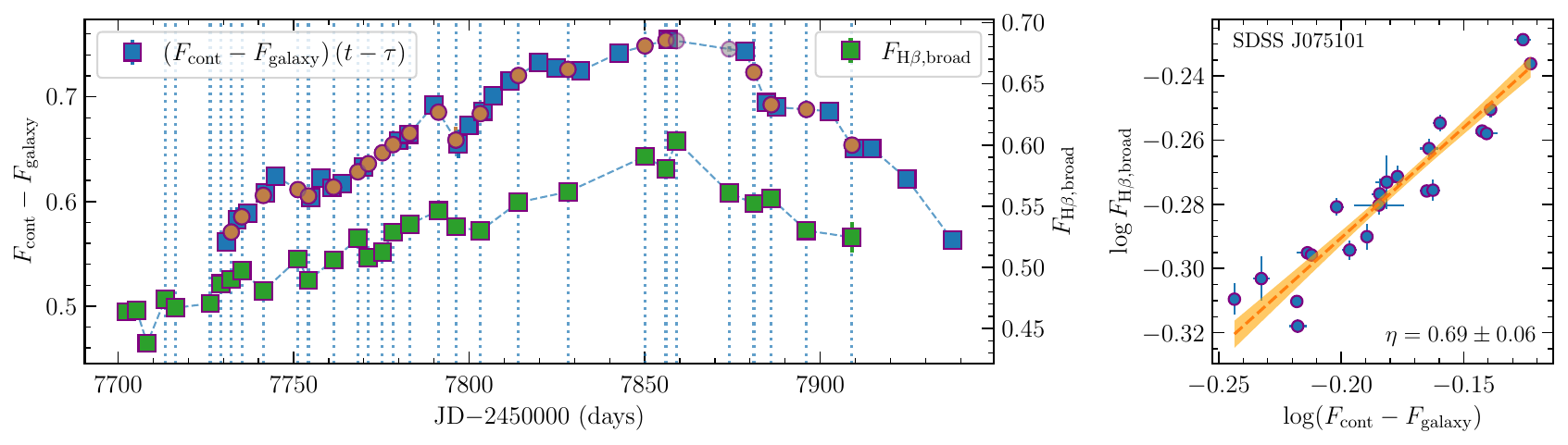}
    \caption{An example of $\eta$ measurement. In the left panel, the green and blue squares represent the broad H$\beta$ light curve and the host-subtracted rest-frame 5100~\AA\ continuum light curve (shifted by the time delay $\tau_{\rm H\beta}$), respectively. The dashed lines connecting the data points indicate their linear interpolations. The brown circles mark the interpolated continuum values at the corresponding epochs of the H$\beta$ points (aligned with vertical dotted lines). Gray circles are excluded due to temporal gaps exceeding the time lag $\tau_{\rm H\beta}$. In the right panel, the blue points illustrate the relationship between the H$\beta$ flux and the corresponding continuum flux, along with the linear regression (orange dashed line and shaded region). The units of $F_{\rm cont}$ are $\rm 10^{-15}\,erg\,s^{-1}\,cm^{-2}\,$\AA$^{-1}$, while the units of $F_{\mathrm{H}\beta, \rm broad}$ are $\rm 10^{-13}\,erg\,s^{-1}\,cm^{-2}$. }
    \label{fig:lc_eta_fit}
\end{figure*}

\subsection{Measurement} \label{sec:measurement}

\subsubsection{Subtraction of Contamination from Host Galaxies and Narrow Lines}
\label{sec:sub_host}

In RM observations, the contamination of continuum flux by starlight from the host galaxy can be significant, particularly in the case of nearby Seyfert galaxies. This is primarily due to the use of long slits with relatively wide widths (approximately 3--5 arcseconds), which are designed to minimize light loss and mitigate calibration issues arising from tracking inaccuracies. Furthermore, the decomposition of the narrow and broad components of H$\beta$ is typically not performed when measuring the H$\beta$ flux in RM studies, as the fluxes of narrow emission lines are generally considered constant, given the kiloparsec-scale sizes of their corresponding emitting regions (the so-called narrow-line regions). Consequently, it is often deemed unnecessary to subtract the narrow components, as their decomposition may introduce additional uncertainties. Therefore, to obtain reliable measurements of responsivity from the RM data, it is essential to carefully remove contamination from the host galaxy and from the narrow component of the H$\beta$ lines (and, in some cases, the [O {\sc iii}] lines).

Fortunately, the contamination from host galaxies in long slits for most RM objects has been estimated and reported in the literature. Among these, many estimates have been derived from high-resolution images obtained from the \textit{Hubble Space Telescope} (HST), utilizing 2D decomposition to distinguish host components from AGN components \citep[e.g.,][]{bentz_low-luminosity_2013, Yang2024}. For objects lacking high-quality HST observations, their host contamination has been estimated through spectral decomposition \citep[e.g.,][]{barth2013, hu_supermassive_2015}. In cases where the spectra suffer from inadequate spectral resolution (e.g., some SEAMBH objects from \citealt{du_supermassive_2015, du_supermassive_2016, Du_Supermassive_2018}), an empirical relationship between the host-galaxy fraction and the total luminosity of AGNs, as established in \cite{Shen2011}, has been employed. Additionally, the narrow H$\beta$ components for most RM objects have also been reported in the literature \citep[e.g.,][]{kaspi2005, bentz_lick_2009, denney_reverberation_2010}. We collected these values and subtracted them from the corresponding continuum and H$\beta$ light curves.

For objects for which host contamination and narrow-line contributions were not available, we determined these values through spectral decomposition using \texttt{DASpec}\footnote{\url{https://github.com/PuDu-Astro/DASpec}} \citep{Du2024}, employing the same methodology as described in \cite{hu_supermassive_2015} and  \cite{Du2019}. This multi-component spectral fitting code is specifically designed for AGNs and is based on the Levenberg-Marquardt algorithm \citep{press1992}, featuring a graphical user interface. The details of the components used in the spectral fitting are provided in the following section.

\begin{figure*}[!t]
    \centering
    \includegraphics[width=\textwidth]{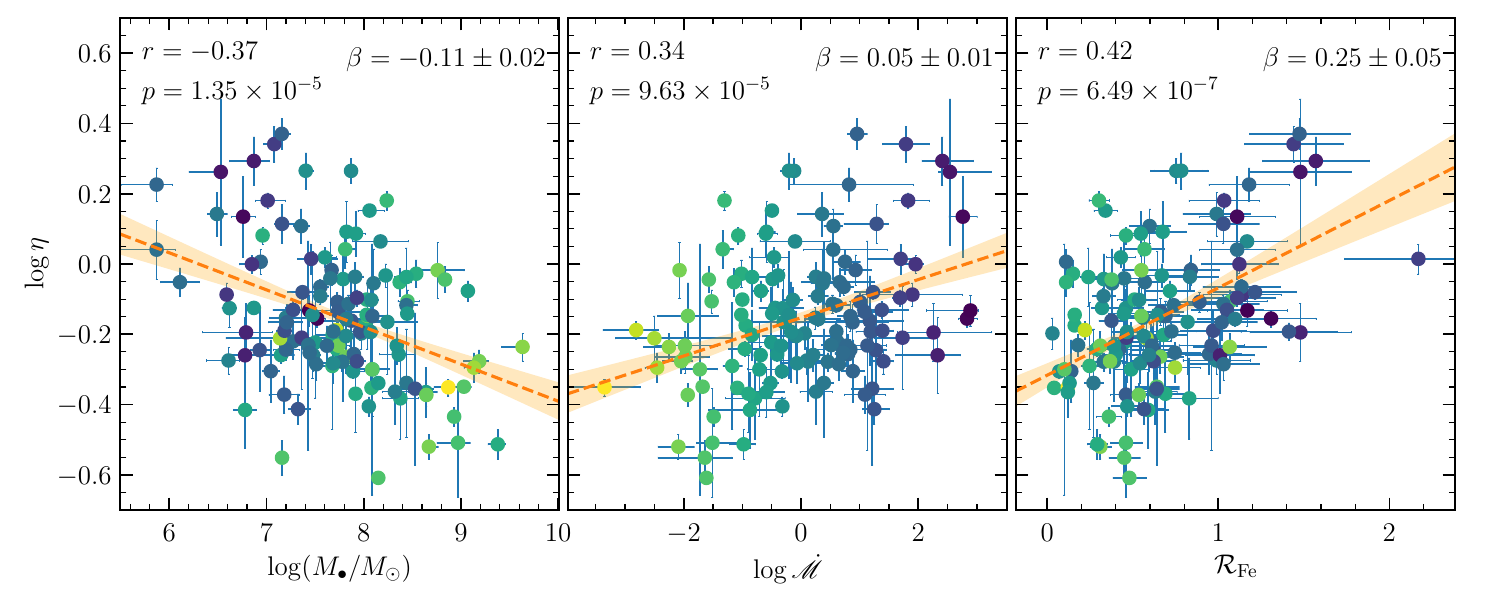}
    \caption{Correlations between responsivity $\eta$ and SMBH mass $M_{\bullet}$, accretion rate $\Mdot$, and iron strength $\RFe$, color-coded by $\Mdot$. Here, $r$ and $p$ denote the Pearson correlation coefficients and their associated null probabilities, respectively, while $\beta$ represents the slope of the linear regressions. }
    \label{fig:logeta_M_RFe}
\end{figure*}

\subsubsection{Spectral Fitting and Strength of Iron Emission}\label{sec:Spectral Fitting}

For objects requiring spectral decomposition, we modeled the observed spectra using the following components:
(1) a power-law continuum; (2) an Fe\,\textsc{ii} template constructed from the narrow-line Seyfert~1 galaxy I~Zw~1 \citep{boroson1992}; (3) a simple stellar population template from \citet{Bruzual2003} to model the host-galaxy starlight; (4) two or more Gaussians for the broad H$\beta$ profile; (5) single Gaussians for broad $\Heii\lambda4686$ and $\Hei\lambda5876$; and (6) one or two Gaussians for each narrow emission line, including H$\beta$, $\OIII\lambda\lambda4959,5007$, $\Heii\lambda4686$, $\Hei\lambda4471$, and coronal lines when required. Further details of the fitting procedure can be found in \citet{hu_supermassive_2015} and \citet{Du2019}.

The relative strength of the optical Fe\,\textsc{ii} emission with respect to the broad H$\beta$ line is defined as $\RFe = F_{\rm Fe\,II}/F_{\rm H\beta,\,broad}$, where $F_{\rm Fe\,II}$ denotes the integrated Fe\,\textsc{ii} flux over 4434--4684~\AA\ and $F_{\rm H\beta,\,broad}$ is the broad H$\beta$ flux. $\RFe$ correlates with the Eddington ratio or accretion rate and can therefore serve as a proxy for either quantity. For objects without published measurements of $\RFe$ \citep[e.g.,][]{Du2019}, the spectral fitting was also used to derive $\RFe$. In addition to the accretion rate, we also present the correlation between $\eta$ and $\RFe$ in the following analysis for comparison.

\subsubsection{Interpolation of Light Curves}
\label{sec:interpolation}

In observations, the broad H$\beta$ line responds to variations in the continuum with a time delay of $\tau_{\rm H\beta}$, owing to the spatial extension of BLRs and the corresponding light-travel time. This time delay is a central objective of the RM campaign and must be accounted for before measuring responsivity. For each pair of continuum and H$\beta$ light curves from a RM campaign, we measured their responsivity as follows: (1) shift the continuum light curve by $\tau_{\rm H\beta}$ days; (2) re-sample the shifted continuum light curve using linear interpolation to match the epochs of the H$\beta$ light curve; and (3) eliminate epochs where the interval between the two adjacent continuum points used for interpolation exceeds $\tau_{\rm H\beta}$ days. Step (3) is implemented to avoid interpolation based on two points with excessive separation, which could introduce bias due to the aliasing effect, thereby ensuring that the interpolation remains as reliable as possible. We also tested that adjusting the criterion in Step (3) to a smaller value (e.g., $0.5 \tau_{\rm H\beta}$) does not significantly affect the results.

\subsubsection{Measurement of Responsivity} \label{sec:Responsivity}

We focus on the responsivity of the H$\beta$ emission line in this work and determine $\eta$ by fitting the relation
\begin{equation}
    \log F_{\rm H\beta,\, broad} = \zeta + \eta \, \log\!\left(F_{\rm cont} - F_{\rm galaxy}\right),
\end{equation}
where $F_{\rm cont}$ denotes the rest-frame 5100~\AA\ continuum flux, $F_{\rm galaxy}$ is the host-galaxy contribution within the spectroscopic aperture (see Section~\ref{sec:sub_host}), $F_{\rm H\beta,\, broad}$ denotes the flux of the broad H$\beta$ line in the light curve after careful subtraction of the narrow-line components, and $\zeta$ is the intercept. The regression was performed using the \texttt{BCES}\footnote{\url{https://github.com/rsnemmen/BCES}} algorithm \citep{Akritas1996_BCES, Nemmen2012_BCES}, which accounts for measurement uncertainties in both variables. We adopted the BCES bisector estimator because both quantities are fluxes and neither can be regarded as the strictly independent variable. An example of the $\eta$ measurement based on the continuum and H$\beta$ light curves is shown in Figure~\ref{fig:lc_eta_fit}.

The effective H$\beta$ responsivity measured here is defined as the response of the H$\beta$ emission-line variability relative to the 5100~\AA\ continuum, after accounting for the continuum--emission-line time delay. We refer to it as ``effective'' because the responsivity is not measured directly relative to the driving ionizing continuum at 1~Ryd, which is not directly observable, but is instead inferred from the variability of the 5100~\AA\ continuum. We assume that the continuum spectral shape does not vary appreciably over the observed variability range for the present sample, such that the variability of the 5100~\AA\ continuum can serve as a proxy for that of the driving continuum at 1~Ryd. For simplicity, we refer to this effective responsivity simply as ``responsivity'' throughout the remainder of this paper.

For objects with multiple RM campaigns, we first measured $\eta$ from each campaign independently and then combined the results using a weighted average. Because some light curves suffer from relatively low signal-to-noise ratios or sparse temporal sampling, the corresponding $\eta$ estimates can carry large uncertainties. To ensure robust measurements, we retained only those campaigns with a Pearson correlation coefficient of $r \geq 0.6$ and a detection significance of $\eta / \eta_{\rm err} \geq 1$, discarding those that do not satisfy these criteria. We have verified that adopting slightly more or less stringent thresholds does not affect the conclusions of this work. Our final sample comprises 128 AGNs with a total of 175 light curves.

\section{Analysis} \label{sec:analysis}

\subsection{Correlations with SMBH Mass, Accretion Rate, and Fe\,{\sc ii} Strength}\label{sec:Correlations}

We examine the correlations of $\eta$ with SMBH mass $M_{\bullet}$, accretion rate $\mathdotM$, and $\RFe$. The RM campaigns are designed to measure the SMBH masses in AGNs. All objects in the current sample have RM measurements, allowing their SMBH masses to be readily obtained from the measured time lags based on the formula
\begin{equation}
\M = f_{\mathrm{BLR}} \frac{\Delta V^2 R_{\mathrm{BLR}}}{G},
\label{equ:BH mass}
\end{equation}
where $G$ is the gravitational constant, $\Delta V$ is the line width (which may be taken as either the FWHM or the line dispersion), and $f_{\mathrm{BLR}}$ is a factor that accounts for the geometry and kinematics of the BLR \citep[e.g.,][]{Onken2004, grier_reverberation_2012, Ho2014, Woo2015, Yang2024}.  The value of the $f_{\mathrm{BLR}}$ factor can be calibrated using the $M_{\bullet}$–$\sigma_{\rm bul}$, $M_{\bullet}$–$L_{\rm bul}$, or $M_{\bullet}$–$M_{\rm bul}$ relations in inactive galaxies, where $\sigma_{\rm bul}$, $L_{\rm bul}$, and $M_{\rm bul}$ represent the stellar velocity dispersion, luminosity, and mass of the galaxy bulge, respectively \citep[e.g.,][]{Onken2004, grier_reverberation_2012, Ho2014, Woo2015, Yang2024}. Alternatively, the $f_{\mathrm{BLR}}$ factor can be directly obtained through dynamical modeling of the BLR for individual AGNs \citep[e.g.,][]{Pancoast2014, Li2018}. The derived value of $f_{\mathrm{BLR}}$ is close to unity if $\Delta V$ is the FWHM, or approximately 5 if $\Delta V$ is the line dispersion (see the aforementioned references). For simplicity, we adopt $f_{\mathrm{BLR}}=1$ to estimate the SMBH mass, while acknowledging the possible scatter of approximately 0.3–0.5 dex in its value for individual objects and its potential dependence on AGN properties (see more details in, e.g., \citealt{Yang2024}).

\begin{figure*}[!t]
    \centering
    \includegraphics[width=\textwidth]{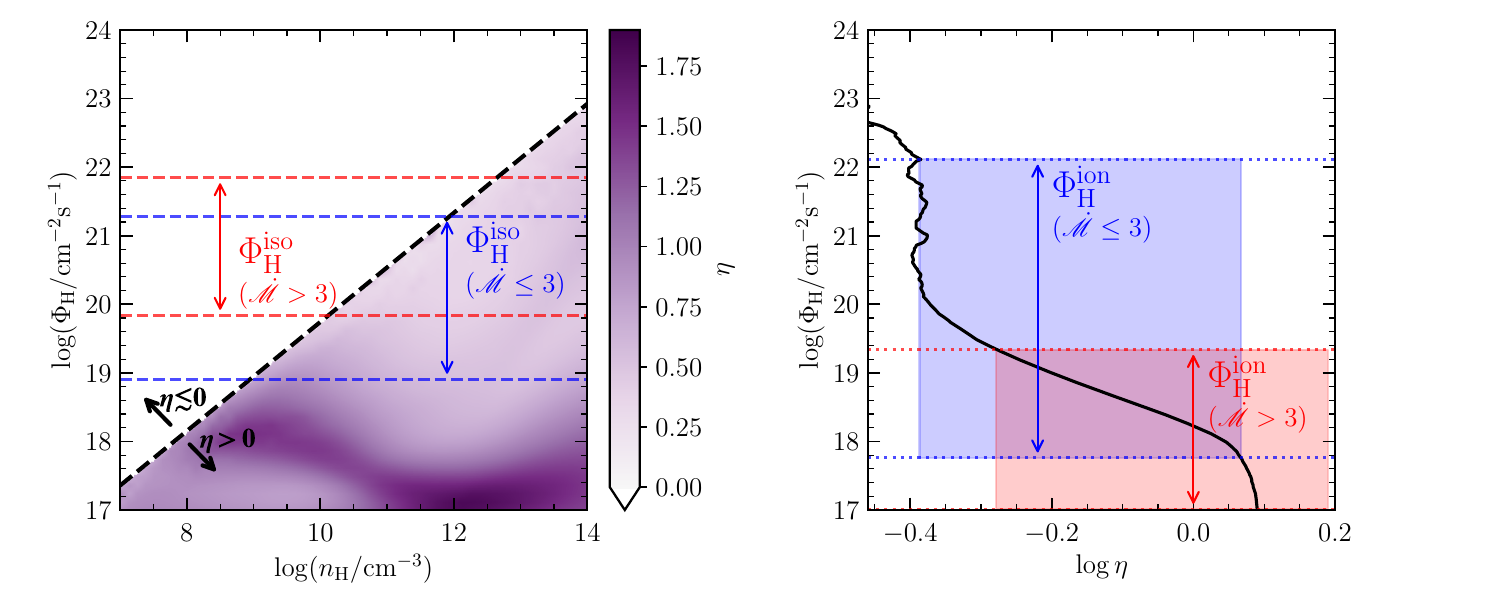}     
    \caption{Left panel: Responsivity $\eta$ as a function of gas number density $\log \nH$ and the surface flux of ionizing photons $\log \PhiH$. The black dashed line roughly separates the regions of positive and negative responsivity. Blue (red) dashed lines indicate the range of isotropic flux $\PhiHiso$ for objects with $\Mdot \leq 3$ ($\Mdot > 3$). Right panel: Relation between the averaged $\eta$ and $\log \PhiH$, demonstrating that responsivity increases as the ionizing photon flux decreases. Blue (red) squares denote the ranges of $\PhiHion$ inferred from the 10th--90th percentile ranges of responsivities $\eta$ for objects with $\Mdot \leq 3$ ($\Mdot > 3$). It is evident that $\PhiHion$ is more consistent with $\PhiHiso$ when $\Mdot \leq 3$; however, they diverge significantly for $\Mdot > 3$. }
    \label{fig:cloudy}
\end{figure*}

For the accretion rate, we adopt the dimensionless form used in \citet{du_supermassive_2014} and \citet{du_supermassive_2016},  
\begin{equation}
\mathdotM = 20.1\left(\frac{\ell_{44}}{\cos i}\right)^{3/2} m_7^{-2},
\label{equ:accretion rate}
\end{equation}
which is derived from the standard accretion disk model \citep{shakura1973}. Here $\ell_{44}$ is the monochromatic luminosity at 5100\,\AA\ in units of $10^{44}\,\mathrm{erg\,s^{-1}}$, $m_{7} = M_{\bullet} / 10^{7} M_{\odot}$ is the SMBH mass in units of $10^{7}\,M_{\odot}$, and $i$ denotes the disk inclination angle. Consistent with previous studies \citep{du_supermassive_2014, du_supermassive_2016, Du_Supermassive_2018}, we adopt a fixed value of $\cos i = 0.75$ (i.e., $i \approx 40^\circ$). 

Figure~\ref{fig:logeta_M_RFe} shows the correlations between responsivity and $M_{\bullet}$, $\mathdotM$, and $\RFe$. The Pearson correlation coefficient $r$ and its associated null probabilities $p$ are shown in the upper-left corner of each panel. It is obvious that the responsivity of broad H$\beta$ decreases when SMBH mass increases, but increases with higher accretion rate. It is also positively correlated with $\RFe$. The small null probabilities ($p < 10^{-4}$) indicate that the three correlations are all significant. We therefore perform linear regressions of the form
\begin{equation} \label{eq:eta}
\log \eta = \alpha + \beta \Xi,
\end{equation}
where $\Xi$ is $\log (M_{\bullet}/M_{\odot})$, $\log \mathdotM$, or $\RFe$, and $\alpha$ and $\beta$ are the intercept and slope, respectively. Because $M_{\bullet}$, $\mathdotM$, and $\RFe$ can be treated as independent variables while $\eta$ serves as the dependent variable, we adopt the \texttt{BCES} Y\,$|$\,X estimator for the regression. The best-fit slopes $\beta$ for the three correlations are reported in the upper-right corner of each panel in Figure~\ref{fig:logeta_M_RFe}.

\subsection{Surface Density of Ionizing Photons}
\label{sec:ionizing photons}

The anisotropy of the ionizing radiation from the accretion disk can be evaluated by comparing the surface density of the ionizing flux on BLR inferred from the observer's line of sight under the assumption of isotropic emission ($\PhiHiso$) with the ionizing flux actually received by the BLR ($\PhiHion$). A substantial discrepancy between $\PhiHiso$ and $\PhiHion$ signifies strong anisotropy, whereas similar values indicate that the radiation field is only weakly anisotropic. We define $\PhiHiso$ and $\PhiHion$ as follows.

\subsubsection{$\PhiHiso$: Ionizing Photons under the Isotropic Assumption} \label{sec:PhiHiso}

If the central ionizing source emits isotropically, the surface density of ionizing photons incident on the BLR clouds is
\begin{equation}
    \PhiHiso = \frac{Q(\mathrm{H})}{4\pi R_{\mathrm{H\beta}}^2}
              = \frac{D_{\rm AGN}^2}{R_{\mathrm{H\beta}}^2} \cdot 
              \frac{F_{\nu, \rm ion}}{h\nu_{\rm ion}},
\label{eq:phi_H_iso}
\end{equation}
where $Q(\mathrm{H})$ is the total ionizing photon production rate, $F_{\nu, \rm ion}$ is the ionizing continuum flux at the hydrogen ionization threshold, and $D_{\rm AGN}$ is the luminosity distance to the AGN. Here, we adopt $h\nu_{\rm ion}=1~{\rm Ryd}$, corresponding to 912~\AA, and use the monochromatic photon flux at the hydrogen ionization threshold as a characteristic estimate of the hydrogen-ionizing photon flux, motivated by the rapid decrease of the hydrogen photoionization cross-section toward higher photon energies. We estimate the continuum flux density at 1~Ryd from the observed monochromatic flux at 5100 \AA, with the relation $F_{912\,\text{\AA}} \approx \kappa\, F_{5100\,\text{\AA}}$, where $\kappa$ is a conversion factor. 

Two cases for the value of $\kappa$ are considered: (1) a constant $\kappa$ for simplicity and (2) a $\kappa$ that depends on the SMBH mass and accretion rate based on an accretion disk model. For the first case, we employ $\kappa = 2.2$, which is derived from the classic spectral energy distribution (SED) of AGNs presented in \cite{Mathews1987}. For the second case, we derive the $\kappa$ for each individual object, as their SEDs may vary significantly with SMBH masses and accretion rates. Here, we obtain the $\kappa$ for sub-Eddington AGNs in the present sample based on the standard accretion disk model \citep{shakura1973}, while for super-Eddington AGNs, we use the code \texttt{Satan}\footnote{\url{https://github.com/LiyrAstroph/Satan}} \citep[][]{Chen2004_slim_disk}, which is based on the slim disk model \citep[e.g.,][]{abramowicz1988, Wang2014_self}. In the second case, we adopt an inner disk radius of $R_{\mathrm{in}} = 6\,R_{\mathrm{g}}$, corresponding to a non-rotating black hole, and an outer radius of $R_{\mathrm{out}} = 10^{4}\,R_{\mathrm{g}}$, where $R_{\mathrm{g}} = GM/c^{2}$ is the gravitational radius. $\M$ and $\Mdot$ are obtained from their RM data (see Section \ref{sec:Correlations}).

\begin{figure*}[!t]
    \centering
    \includegraphics[width=\textwidth]{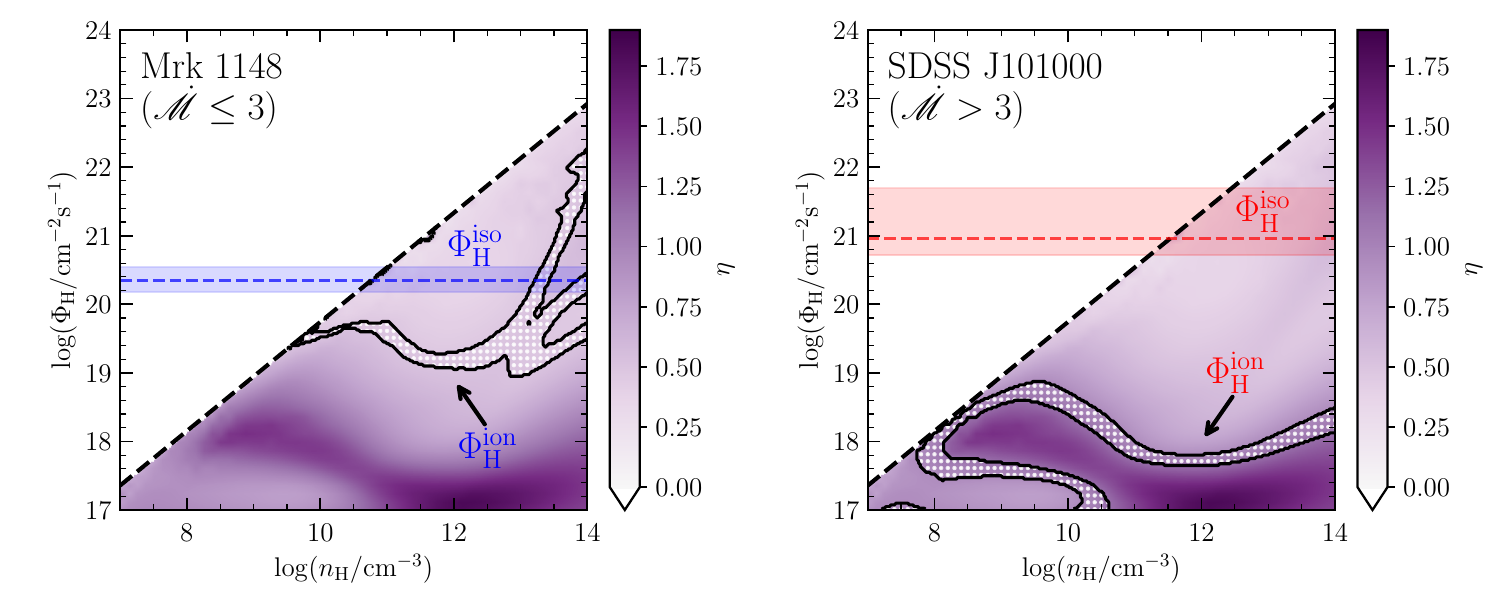}
    \caption{
    Examples of measurements of the ionizing photon flux incident on the BLR clouds, $\PhiHion$, inferred from the H$\beta$ responsivity under the single-cloud approximation (black‑outlined, white‑dotted region), compared with the ionizing flux inferred along the observer’s line of sight assuming isotropic emission, $\PhiHiso$ (dashed line with shading). Left and right panels show objects with low and high accretion rates, respectively. 
    }   
    \label{fig:cloudy_examples}
\end{figure*}

\subsubsection{$\PhiHion$: Ionizing Photons incident on BLR clouds} \label{sec:PhiHion}

Here, we perform a simple photoionization calculation using \texttt{CLOUDY} v17.02 \citep{ferland2017} to derive the surface flux of ionizing photons from the responsivity. Following previous photoionization calculations of BLRs \citep[e.g.,][]{korista2000, Korista_Goad_2004, Du2023_cloudy}, we generated a grid of models spanning gas number densities of $7 \leq \log (\nH/{\rm cm^{-3}}) \leq 14$ and surface fluxes of ionizing photons of $17 \leq \log (\PhiH/{\rm cm^{-2}\ s^{-1}}) \leq 24$, with a step size of 0.125 dex in both parameters. We assumed a slab geometry with a column density of $10^{23}\ {\rm cm^{-2}}$ for the line-emitting entities. The characteristic SED from \cite{Mathews1987} was utilized as the ionizing source, along with a metallicity of solar abundance. For comparison, we also adopted the SED applicable to super-Eddington sources from \cite{ferland2020}, using a metallicity of $5 Z_{\odot}$, as high-accretion-rate AGNs are generally believed to possess relatively high metallicities \citep[e.g.,][]{shemmer2004, panda2019, sniegowska2021}, where $Z_{\odot}$ represents solar metallicity. 

The responsivity for various parameters of $\nH$ and $\PhiH$ is presented in the left panel of Figure~\ref{fig:cloudy} (see \citealt{Korista_Goad_2004} and \citealt{Du2023_cloudy} for details). In the upper-left half of the parameter space shown in the left panel of Figure~\ref{fig:cloudy}, the responsivity is generally negative (more details can be found in \citealt{Du2023_cloudy}). Therefore, this region falls outside the scope of the present study because all AGNs in this sample exhibit a positive response. We thus mask and leave blank this region in the figure. Our focus is on the lower-right portion of the parameter space, as separated by the dashed black line in this panel. In the region of positive responsivity, it is evident that the responsivity is generally higher in the low $\PhiH$ state ($17 \leq \log (\PhiH/{\rm cm^{-2}\ s^{-1}}) \leq 19$; \citealt{Korista_Goad_2004}). In this regime, the clouds in the BLR are more sensitive to changes in the ionizing continuum. In contrast, at high $\PhiH$ ($19 \leq \log (\PhiH/{\rm cm^{-2}\ s^{-1}}) \leq 24$), the clouds in the BLR are ionized to a greater extent, resulting in a lower responsivity to the ionizing photons. Consequently, the surface flux of ionizing photons received by the BLR clouds ($\PhiHion$) can be estimated from the responsivity based on the \texttt{CLOUDY} calculations.

In \texttt{CLOUDY}, $\PhiH$ is defined as the total surface flux of hydrogen-ionizing photons above the hydrogen ionization threshold of 1~Ryd, which differs from the monochromatic photon flux at 1~Ryd defined in Eqn.~(\ref{eq:phi_H_iso}). Therefore, before the subsequent analysis, we convert the $\PhiH$ values obtained from \texttt{CLOUDY} to the corresponding monochromatic photon fluxes at 1~Ryd to ensure a consistent definition of $\PhiH$ throughout our analysis.

\begin{figure*}[!ht]
    \centering
    \includegraphics[width=\textwidth]{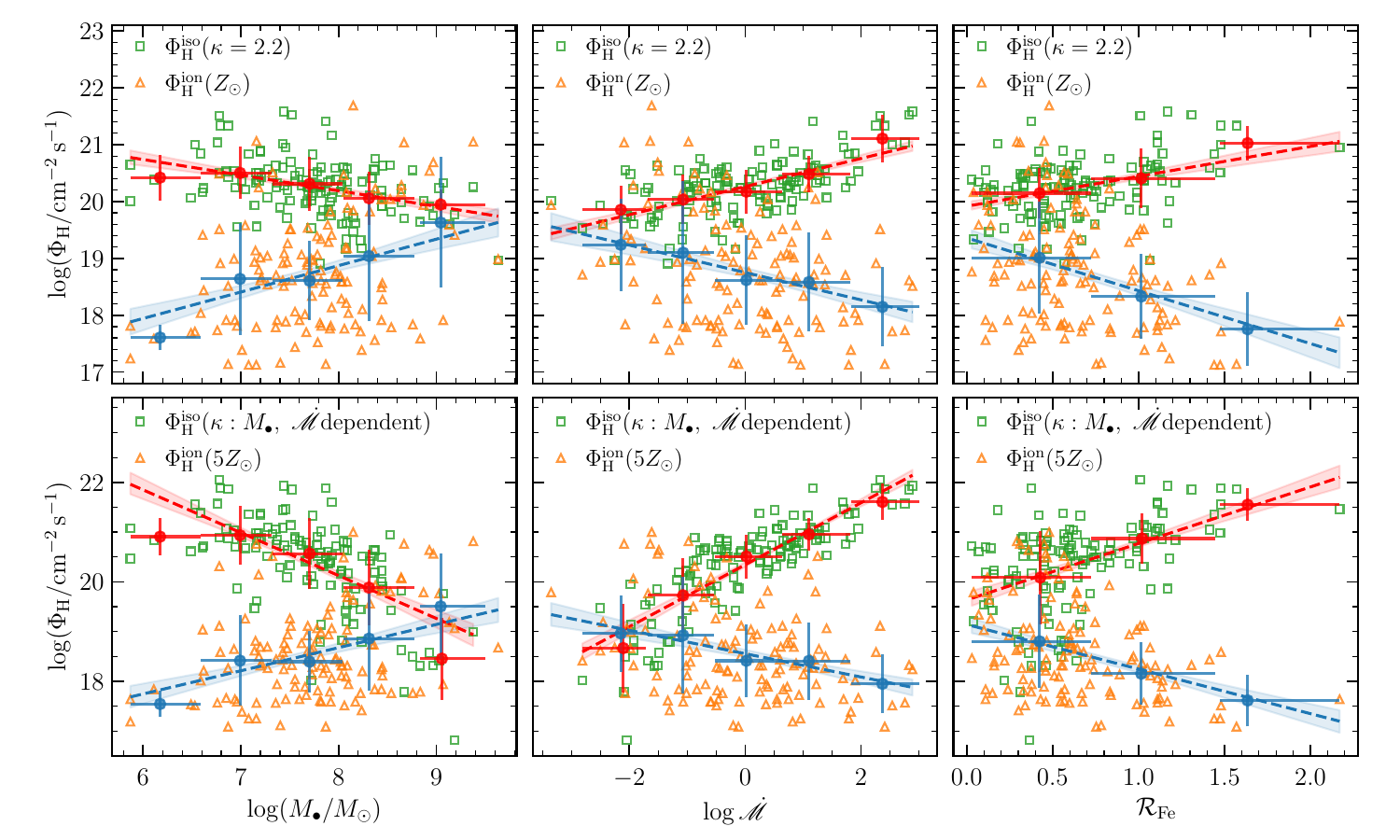}
    \caption{Comparisons between $\PhiHion$ (orange triangles), inferred from the H$\beta$ responsivity under the single-cloud approximation, and $\PhiHiso$ (green squares), inferred along the observer's line of sight assuming isotropic emission, along with their correlations with $\log (\M/M_{\odot})$, $\log \Mdot$, and $\RFe$. The upper panels present results based on the scenario with a constant $\kappa$, utilizing \texttt{CLOUDY} calculations at solar metallicity with the classic SED \citep{Mathews1987}. The lower panels display results from the scenario where $\kappa$ depends on $M_{\bullet}$ and $\Mdot$, using \texttt{CLOUDY} calculations for a metallicity of $5Z_\odot$ and a super-Eddington SED \citep{ferland2020}. The weighted means of $\PhiHion$ and $\PhiHiso$ are indicated as blue and red points, respectively, while the dashed lines represent the corresponding linear regressions. }
    \label{fig:phiH_accretion_rate}
\end{figure*}

\subsection{A Simple Comparison between $\PhiHiso$ and $\PhiHion$}
\label{sec: PhiHiso and PhiHion}

We can first compare the $\PhiHiso$ and $\PhiHion$ of sub- and super-Eddington AGNs in the present sample. Following \cite{du_supermassive_2016}, here we define the super-Eddington AGNs using the conservative criterion of $\mathdotM \geq 3$, considering the lowest possible radiation efficiency. The objects that do not meet this requirement are sub-Eddington AGNs. The range of $\PhiHiso$ as determined from Eqn.~(\ref{eq:phi_H_iso}) for sub- and super-Eddington AGNs are plotted overlaying the responsivity calculated from \texttt{CLOUDY} in the left panel of Figure~\ref{fig:cloudy}, marked by the blue and red dashed lines, corresponding to $18.9 < \log \PhiHiso < 21.3$ and $19.8 < \log \PhiHiso < 21.8$, respectively. Here, we present only the results based on the scenario with a constant $\kappa$, utilizing \texttt{CLOUDY} calculations at solar metallicity with the classic SED \citep{Mathews1987}. The results from the scenario where $\kappa$ depends on $\M$ and $\Mdot$, using \texttt{CLOUDY} calculations for a metallicity of $5Z_\odot$ and a super-Eddington SED \citep{ferland2020}, are provided in the Appendix \ref{sec:appendix 5z}. Under the assumption of isotropic emission from the accretion disk, the ionizing photon flux inferred from the continuum is generally similar for AGNs with low and high accretion rates, with $\PhiHiso$ for high-$\Mdot$ objects being marginally higher. Consequently, the corresponding range of responsivity from \texttt{CLOUDY} that $\PhiHiso$ covers for sub- and super-Eddington AGNs are also similar. However, this finding is obviously not consistent with the fact that the responsivity of super-Eddington AGNs are systematically higher than those of sub-Eddington AGNs (see Figure~\ref{fig:logeta_M_RFe}, middle panel). Thus, this implies that the $\PhiHiso$ may not represent the ionizing photons incident on the BLR clouds. 

As aforementioned, from the photoionization calculation, the responsivity of the clouds with lower $\PhiH$ is higher than the case with high $\PhiH$. We plot the responsivity $\eta$ as a function of ionizing flux $\PhiH$ as a black curve in the right panel by averaging the responsivity of all $\nH$ at each fixed $\PhiH$ in the positive responsivity regime in the left panel of Figure~\ref{fig:cloudy}. Additionally, the 10th--90th percentile ranges of the $\eta$ distributions for the sub- and super-Eddington AGN subsamples are plotted as blue and red shaded regions, respectively, in the right-hand panel of Figure~\ref{fig:cloudy}. The upper and lower bounds of $\PhiHion$, determined from the intersections between the observed ranges of $\eta$ and the $\eta(\PhiH)$ relation, are marked by blue and red dashed lines for sub- and super-Eddington AGNs, respectively. Because of the higher $\eta$ in super-Eddington AGNs, their $\PhiHion$ are much lower than the values of sub-Eddington AGNs, which are very different from the case of $\PhiHiso$ where the values are more similar for sub- and super-Eddington AGNs. $\PhiHion$ in sub-Eddington AGNs is generally consistent with their $\PhiHiso$. By contrast, $\PhiHion$ of the super-Eddington objects deviates significantly from their $\PhiHiso$. Thus, the ionizing photon flux that the BLR can see is lower than that inferred under the isotropic assumption from the observed continuum for super-Eddington AGNs. 

As a test, we also recomputed the averaged $\eta(\PhiH)$ relation after conservatively excluding the low-$\PhiH$, high-$\nH$ region with $\log Uc<4.5$, where the gas is weakly ionized and the dust-free \texttt{CLOUDY} calculations may be less appropriate \citep[see][]{Korista_Goad_2004}. Here $U$ is the ionization parameter. The resulting $\eta(\PhiH)$ relation is very similar to that obtained without this additional cut, and the conclusion that the H$\beta$ responsivity increases toward lower incident ionizing photon flux remains unchanged. In particular, the $\PhiHion$ region inferred from the observed responsivities remains clearly offset from the $\PhiHiso$ region.

\subsection{Comparison Between $\PhiHiso$ and $\PhiHion$ Based on Single-Cloud Approximation}
\label{sec:PhiHion_single_cloud}

\subsubsection{Correlations Between $\PhiHiso$, $\PhiHion$, and Accretion Rate}
\label{sec:Correlations between Phi_H}

\begin{figure*}[!t]
    \centering
    \includegraphics[width=\textwidth]{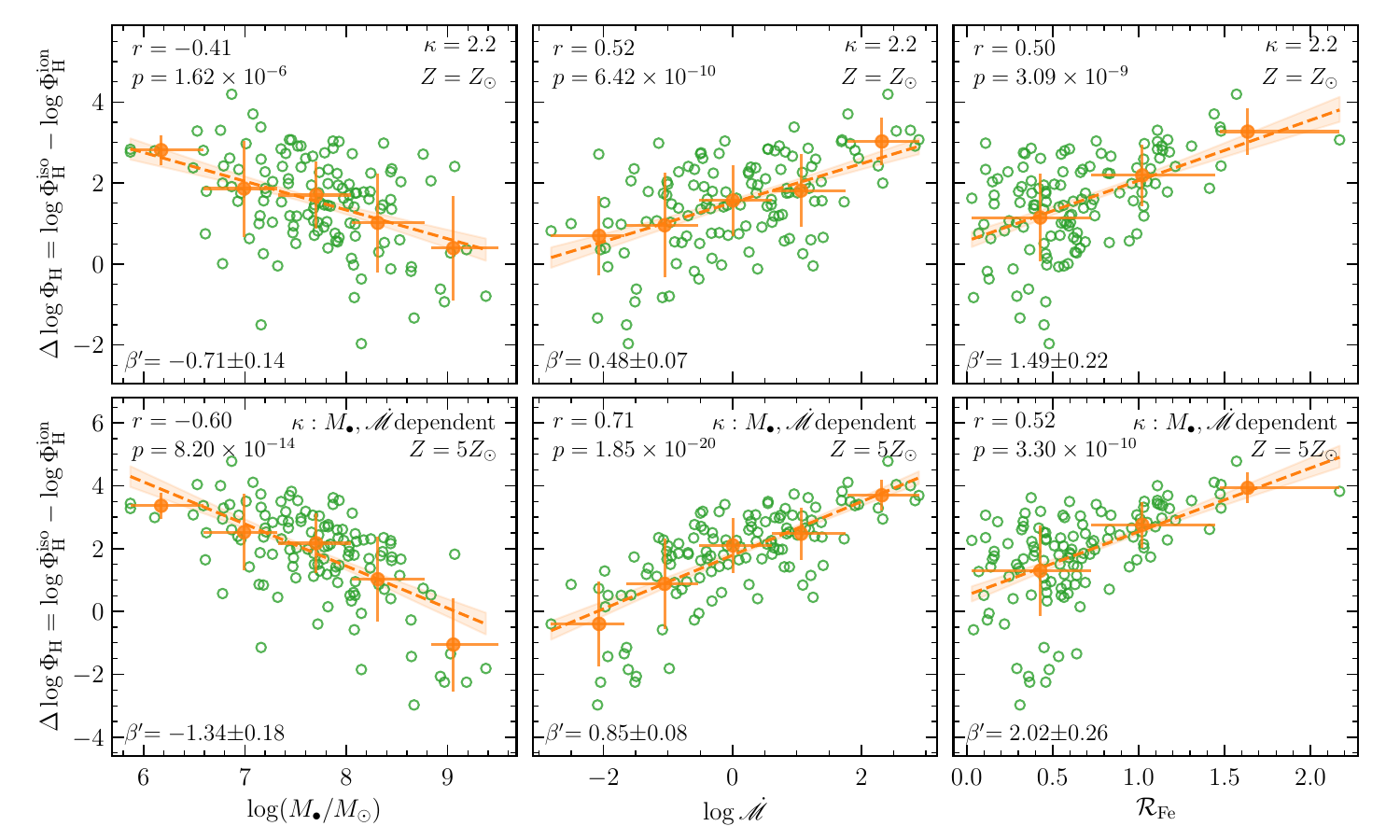}
    \caption{Correlations between $\DeltaPhiH$ (under the single-cloud approximation) and $\log (\M/M_{\odot})$, $\log \Mdot$, and $\RFe$. The weighted means are indicated as orange points, while the dashed lines represent the corresponding linear regressions. The values of $r$ and $p$ denote the Pearson correlation coefficients and their associated null probabilities, respectively, and $\beta^{\prime}$ represents the slope of the regression. The upper and lower panels correspond to the two scenarios as shown in Figure~\ref{fig:phiH_accretion_rate}. }
    \label{fig:phiH_accretion_rate_delta}
\end{figure*}

We can measure the $\PhiHion$ of each individual object from its responsivity based on the \texttt{CLOUDY} responsivity result under the single-cloud approximation. In this approximation, the BLR is represented by a single cloud characterized by a specific incident ionizing photon flux, gas density, and responsivity. Figure~\ref{fig:cloudy_examples} illustrates two examples of $\PhiHion$ measurements for individual objects, assuming solar metallicity and the SED from \cite{Mathews1987}. Examples corresponding to a metallicity of $5Z_\odot$ and the SED of \cite{ferland2020} are presented in Appendix~\ref{sec:appendix 5z}. The regions shaded with white dots and surrounded with black lines are the parameter space of $\nH$ and $\PhiH$ where the responsivity is within the specific value derived from the RM light curves (Section \ref{sec:Responsivity}). The left panel is a sub-Eddington object (Mrk~1148, see \citealt[][]{bao_monitoring_2022}) whereas the right panel is a super-Eddington one (SDSS~J101000, see \citealt[][]{Du_Supermassive_2018}). For comparison, the $\PhiHiso$ obtained from their luminosities and BLR radii (Section \ref{sec:PhiHiso}) are shown overlaid with blue and red dashed lines, with the shaded regions in both panels indicating their measurement uncertainties. From Figure~\ref{fig:cloudy_examples}, it is clear that the ranges of $\PhiHiso$ and $\PhiHion$ are more similar (with significant overlap) in the sub-Eddington AGN while $\PhiHiso$ substantially exceed $\PhiHion$ in super-Eddington AGN.

Such a difference is not limited to these individual cases. In Figure~\ref{fig:phiH_accretion_rate}, we present the correlations between $\PhiHiso$, $\PhiHion$, and the parameters $M_{\bullet}$, $\Mdot$, and $\RFe$ for two scenarios: one with a constant $\kappa$ (for $\PhiHiso$) and based on the \texttt{CLOUDY} calculations at solar metallicity and the classic SED (for $\PhiHion$), and the other with a $\kappa$ that depends on $M_{\bullet}$ and $\Mdot$, using \texttt{CLOUDY} for a metallicity of $5Z_\odot$ and a super-Eddington SED. Three outliers with $\log\PhiHiso < 16$, corresponding to the lowest $\Mdot$ and the largest $\M$ in the sample, are excluded from Figure~\ref{fig:phiH_accretion_rate} and from the subsequent analysis. Their accretion states may fall within or close to the advection-dominated accretion flow regime \citep[e.g.,][]{Narayan1994}, for which the determination of $\kappa$ is subject to large uncertainties under the procedures adopted in Section~\ref{sec:PhiHiso}. From Figure~\ref{fig:phiH_accretion_rate}, it is evident that $\PhiHiso$ decreases with increasing $M_{\bullet}$ but increases with $\Mdot$ and $\RFe$. In contrast, $\PhiHion$ exhibits a distinctly different trend, increasing with $\M$ but decreasing with $\Mdot$ and $\RFe$. Consequently, the discrepancy between $\PhiHiso$ and $\PhiHion$ becomes increasingly pronounced as SMBH mass decreases and accretion rate increases, suggesting that radiation anisotropy is stronger in super-Eddington AGNs and relatively weaker in sub-Eddington AGNs.

\begin{figure*}[!t]
    \centering
    \includegraphics[width=\textwidth]{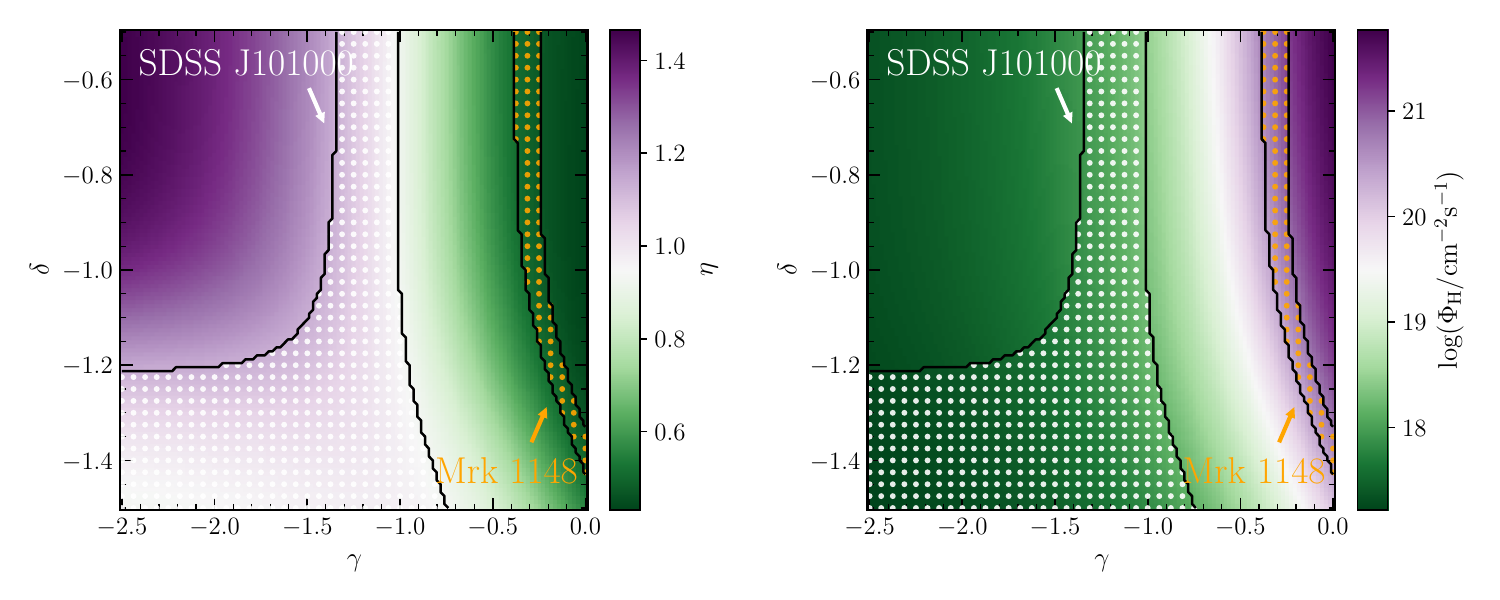}
    \caption{Examples of the allowed LOC parameter space for Mrk~1148 (dashed line with orange shading) and SDSS~J101000 (dashed line with white shading). The color maps show the modeled H$\beta$ responsivity $\eta$ (left) and the corresponding $\log \Phi_{\rm H}$ (right) as functions of the ionizing flux and density weighting indices $\gamma$ and $\delta$, respectively. }
    \label{fig:gamma_delta_example_EW}
\end{figure*}

\subsubsection{Dependence of the Difference Between $\PhiHiso$ and $\PhiHion$ on AGN Properties}

To further investigate the dependence of anisotropy on AGN properties, we define the difference between $\PhiHiso$ and $\PhiHion$ as a quantitative measure of anisotropy:
\begin{equation}
    \Delta\log\PhiH = \log \PhiHiso - \log \PhiHion .
\end{equation}
The correlations of $\Delta\log\PhiH$ with $\M$, $\Mdot$, and $\RFe$ are shown in Figure~\ref{fig:phiH_accretion_rate_delta}. This figure demonstrates that $\Delta\log\PhiH$ exhibits significant positive correlations with both $\log\Mdot$ and $\RFe$, while showing a negative correlation with $\log(\M/M_{\odot})$. The Pearson correlation coefficients and the corresponding null probabilities are indicated in the upper-left corner of each panel. All null probabilities are much smaller than $10^{-5}$, confirming the high statistical significance of these correlations.

Following Eqn. (\ref{eq:eta}), we perform linear regressions of $\Delta\log\PhiH$ against $\M$, $\Mdot$, and $\RFe$ to further quantify these dependencies. The regression is expressed as
\begin{equation} \label{eq:DeltaPhi}
\Delta\log\PhiH = \alpha^{\prime} + \beta^{\prime} \Xi ,
\end{equation}
where $\Xi$ represents $\log(M_{\bullet}/M_{\odot})$, $\log\Mdot$, or $\RFe$, and $\alpha^{\prime}$ and $\beta^{\prime}$ are the intercept and slope, respectively. The best-fit slopes $\beta^{\prime}$ are displayed in the lower-left corners of the panels in Figure~\ref{fig:phiH_accretion_rate_delta}.

$\Delta\log\PhiH$ shows statistically significant correlations with both $\M$ and $\Mdot$ (as well as with $\RFe$, which serves as a proxy for $\Mdot$). However, based on the null probabilities, the dependence on $\Mdot$ is the most significant. On the one hand, both $\M$ and $\Mdot$ are derived from time delays measured in RM data. On the other hand, the $\Mdot$ distributions differ among objects with different $\M$. Consequently, $\M$ and $\Mdot$ are not independent, which may account for the presence of significant correlations with both parameters.

To assess whether $\M$ or $\Mdot$ is the primary driver of anisotropy, we conduct a partial correlation analysis of $\Delta\log\PhiH$ with respect to $\M$ and $\Mdot$, as summarized in Table~\ref{table:partial correlation}. The partial correlation coefficients show that the correlation with $\Mdot$ when controlling for $\M$ is substantially stronger than the correlation with $\M$ when controlling for $\Mdot$. This result indicates that $\Mdot$ is the dominant parameter governing the variation in anisotropy, and supports the \emph{self-shadowing effect} of slim accretion disks proposed by \cite{Wang2014_self}.

\begin{figure*}[!t]
    \centering
    \includegraphics[width=\textwidth]{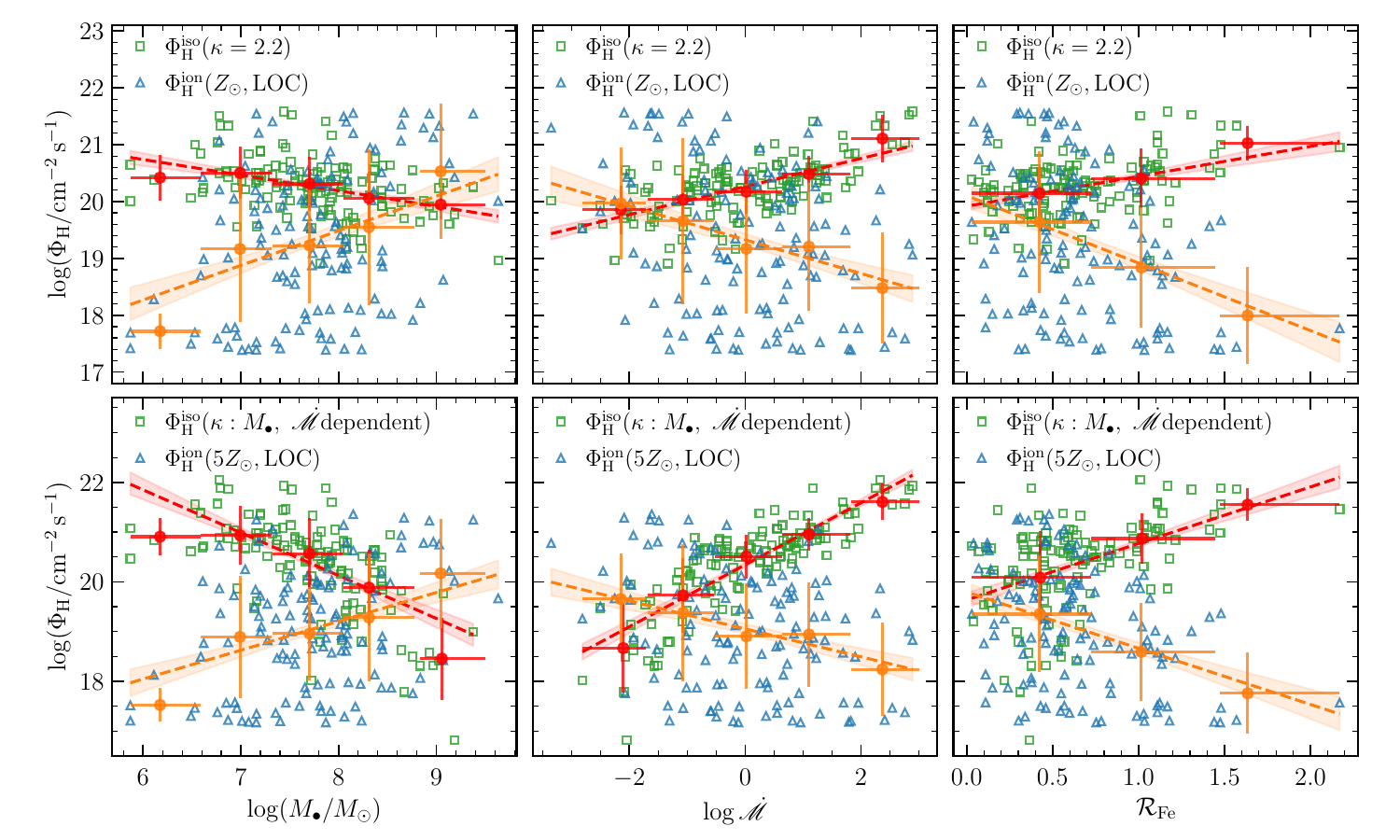}
    \caption{
    Same as Figure~\ref{fig:phiH_accretion_rate}, but for $\PhiHion$ (LOC-based). Blue triangles show the individual $\PhiHion$ estimates obtained in the LOC framework described in Section~\ref{sec:phihion_eta_LOC}, while orange points indicate the weighted means. The upper and lower panels correspond to $Z_{\odot}$ and $5Z_{\odot}$, respectively. }
    \label{fig:phiH_accretion_rate_LOC_EW}
\end{figure*}

\subsection{Comparison Between $\PhiHiso$ and $\PhiHion$ Based on Locally Optimally Emitting Cloud (LOC) Model}
\label{sec:phihion_eta_LOC}

The conversion from the observed H$\beta$ responsivity to the incident ionizing photon flux in Section~\ref{sec:PhiHion_single_cloud} is based on the single-cloud approximation. In reality, however, the BLR is likely to comprise an ensemble of clouds spanning a broad range of ionizing photon fluxes and gas densities. We therefore extend our analysis using the locally optimally emitting cloud (LOC) framework \citep{Baldwin1995, korista2000, Bottorff2002}, which accounts for the distribution of BLR clouds over a range of physical conditions.

Following \citet{Bottorff2002}, the integrated H$\beta$ intensity can be expressed as the sum of the contributions from clouds over the range of physical conditions,
\begin{equation}
I_{\mathrm{H}\beta}\propto\iint  \frac{F_{\mathrm{H}\beta}(\nH, \PhiH)}{\PhiH}
\frac{\partial^2C_{\rm f}}{\partial \nH \partial \PhiH} \, d\nH \, d\PhiH ,
\end{equation}
where ${\partial^2 C_{\rm f}}/{\partial \nH \partial \PhiH}$ is the distribution of BLR clouds over $\nH$--$\PhiH$ parameter space, and $F_{\mathrm{H}\beta}(\nH, \PhiH)$ is the H$\beta$ surface flux emitted by a BLR cloud with hydrogen density $\nH$ and incident ionizing photon flux $\PhiH$. We assume that ${\partial^2 C_{\rm f}}/{\partial \nH \partial \PhiH}$ is separable into ionizing-flux and gas-density distributions ($f(\PhiH)$ and $g(\nH)$), and adopt power-law forms for both,
\begin{equation}
\frac{\partial^2 C_{\rm f}}{\partial \nH \partial \PhiH}
\propto f(\PhiH)g(\nH)
\propto \PhiH^{\gamma} \nH^{\delta}.
\label{eq:R_gamma_nH_delta}
\end{equation}
Under this cloud distribution, the responsivity of the integrated H$\beta$ emission can be written as
\begin{equation}
\eta(\gamma, \delta)=\frac{\displaystyle\iint
\eta(\nH,\PhiH) W_{\eta} \, d\log \nH \, d\log\PhiH}
{\displaystyle\iint W_{\eta} \, d\log \nH \, d\log\PhiH},
\label{eq:eta_LOC}
\end{equation}
where $W_{\eta}=F_{\mathrm{H}\beta}(\nH,\Phi_{\rm H}) \Phi_{\rm H}^\gamma \nH^{\delta+1}$. Thus, $\eta(\gamma,\delta)$ represents the emissivity-weighted average of the individual-cloud responsivities over the adopted LOC cloud distribution.

Previous LOC studies have typically adopted or inferred cloud-distribution indices of $\gamma$ and $\delta$ around $-1$ \citep[e.g.,][]{korista2000, Bottorff2002}. Here, we explore the broader parameter ranges of $-2.5\leq\gamma\leq0$ and $-1.5\leq\delta\leq -0.5$, allowing for a wider range of cloud distributions and encompassing the observed range of H$\beta$ responsivities. In the LOC calculations, we restrict the integration to the region of the $(\nH,\PhiH)$ parameter space with positive responsivity, corresponding to the region shown in Figure~\ref{fig:cloudy}.

\begin{figure*}[!t]
    \centering
    \includegraphics[width=\textwidth]{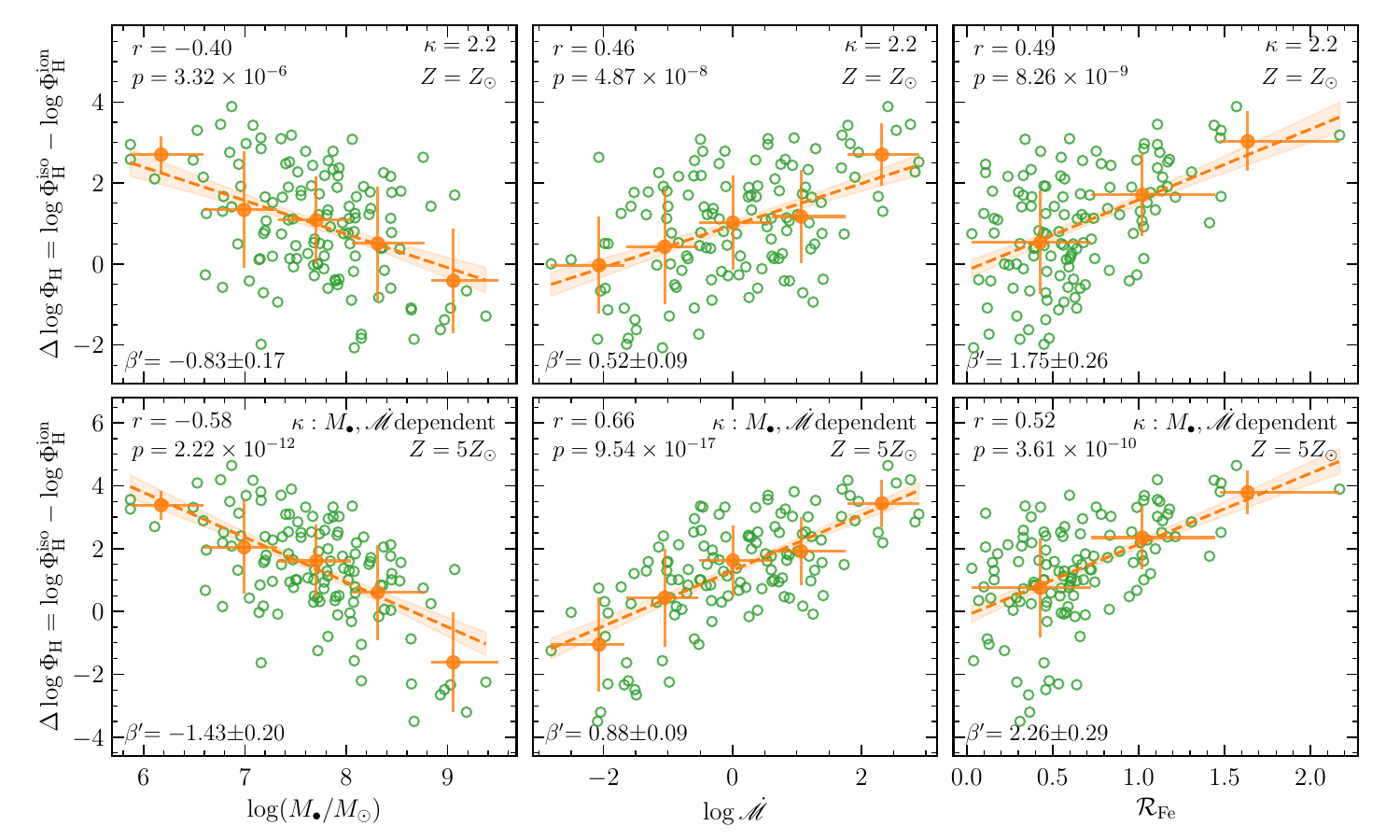}
    \caption{
    Same as Figure~\ref{fig:phiH_accretion_rate_delta}, but for $\DeltaPhiH$ (LOC-based). The upper and lower panels correspond to the two scenarios as shown in Figure~\ref{fig:phiH_accretion_rate_LOC_EW}. The legend and symbols have the same meanings as in Figure~\ref{fig:phiH_accretion_rate_delta}.}
    \label{fig:LOC_phi_H_all_delta_EW}
\end{figure*}

For each source, following the procedure adopted for the single-cloud approximation in Section~\ref{sec:Correlations between Phi_H}, we first constrain the LOC cloud-distribution parameters $(\gamma,\delta)$ using the observed H$\beta$ responsivity. Specifically, we identify the regions in the $(\gamma,\delta)$ parameter space for which the model responsivity $\eta(\gamma,\delta)$, calculated from the LOC calculation (Eqn.~\ref{eq:eta_LOC}), is consistent with the observed responsivity within its uncertainty. For each allowed $(\gamma,\delta)$ pair, the characteristic ionizing photon flux of the BLR clouds is quantified by the emissivity-weighted mean,
\begin{equation}
\log \PhiH(\gamma, \delta) =
\frac{\displaystyle\iint \log\PhiH\,W_{\eta}\,
d\log\nH\,d\log\PhiH}
{\displaystyle\iint W_{\eta}\,
d\log\nH\,d\log\PhiH}.
\label{eq:phi_LOC_EW}
\end{equation}
We then define the LOC-based $\log\PhiHion$ for each source as the median of $\log\PhiH(\gamma,\delta)$ over the allowed region of the $(\gamma,\delta)$ parameter space. Two examples of the resulting allowed $(\gamma,\delta)$ parameter space are shown in Figure~\ref{fig:gamma_delta_example_EW}. The resulting relations between $\PhiHion$ and $M_{\bullet}$, $\dot{\mathcal{M}}$, and $R_{\rm Fe}$ within the LOC framework are shown in Figure~\ref{fig:phiH_accretion_rate_LOC_EW}, while those for $\Delta\log\Phi_{\rm H}$ are presented in Figure~\ref{fig:LOC_phi_H_all_delta_EW}.

The inferred $\PhiHion$ and its relations with $M_{\bullet}$, $\dot{\mathcal{M}}$, and $R_{\rm Fe}$ are remarkably consistent with those obtained under the single-cloud approximation (Figure~\ref{fig:phiH_accretion_rate}). In particular, the increasing discrepancy between $\PhiHiso$ and $\PhiHion$ with increasing accretion rate is preserved in the LOC framework. This consistency demonstrates that our results are robust against the choice between the single-cloud approximation and a more realistic LOC cloud distribution.

As a further check, we also evaluate $\PhiHion$ using the ionizing photon flux at the responsivity-weighted radius, $R_{\rm RW}$, defined by \citet{Bottorff2002}, and examine whether this alternative definition affects the inferred correlations (see Appendix~\ref{sec:phihiso_RRW} for details). The conclusions remain unchanged, indicating that our main results are insensitive to the specific prescription adopted to characterize the characteristic $\PhiHion$ within the LOC framework.

In addition, the $\eta(\PhiH)$ relation obtained from the LOC calculations is broadly consistent with that derived using the simple averaging scheme adopted in Section~\ref{sec: PhiHiso and PhiHion} (see Appendix~\ref{sec:appendix LOC-weighted responsivity} for details). These tests demonstrate that our main conclusions are robust against different treatments of the cloud distribution and the weighting of contributions from clouds with different physical conditions. 

Finally, the potentially different distributions of BLR clouds in the $(\nH,\PhiH)$ plane between sub- and super-Eddington AGNs, as characterized by their different $(\gamma,\delta)$ values, may lead to differences in the reverberation responses of emission lines with different ionization states. In particular, the cloud distributions must also be consistent with the observed ionization stratification of the BLR, which can provide additional constraints on the allowed $(\gamma,\delta)$ parameter space \citep[e.g.,][]{korista2000}. Future theoretical modeling and multi-line RM observations will be valuable for investigating these effects.

\section{Discussion}
\label{sec:discussion}

\begin{figure*}[!t]
    \centering
    \includegraphics[width=0.7\textwidth]{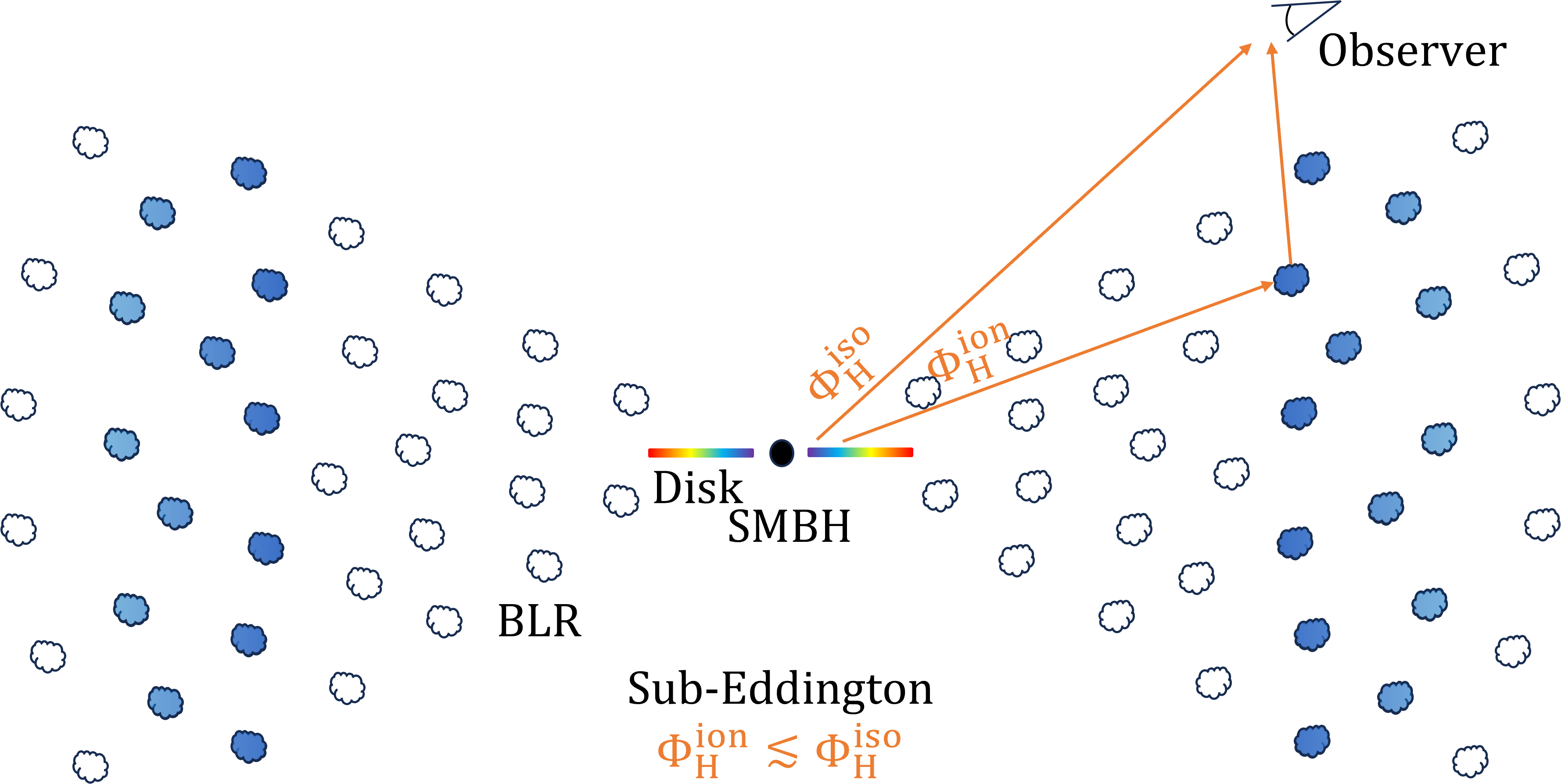}\\ \vspace{0.5cm}
    \includegraphics[width=0.7\textwidth]{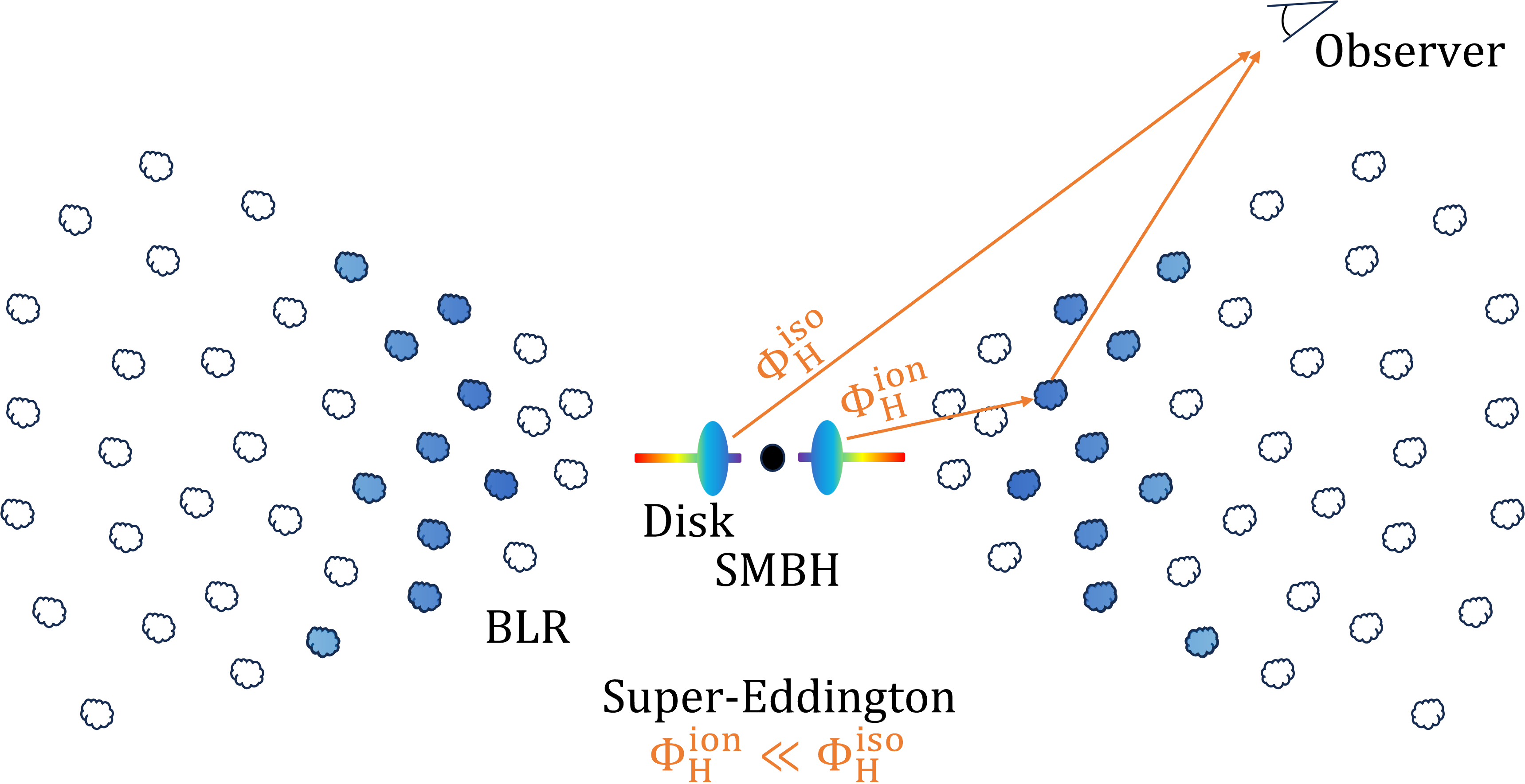}
    \caption{Cartoons illustrating the BLRs illuminated by accretion disks in sub- and super-Eddington AGNs. The blue clouds represent the BLR clouds that respond to continuum variability. In the top panel, for sub-Eddington AGNs with geometrically thin accretion disks, the ionizing photon flux received by the clouds, $\PhiHion$, is comparable to, or only slightly weaker than, the flux inferred under the isotropic assumption, $\PhiHiso$. In the bottom panel, for super-Eddington AGNs, the inner regions of slim disks are geometrically thick, producing anisotropic ionizing radiation such that $\PhiHion \ll \PhiHiso$ for clouds near the equatorial plane, resulting in a smaller effective response radius. The panels are schematic and not to scale.}
    \label{fig:cartoon}
\end{figure*}

\subsection{Anisotropic Radiation in sub-Eddington AGNs}
\label{sec:anisotropy_in_sub_Eddington}

It has been clearly demonstrated that radiation anisotropy is stronger in super-Eddington AGNs. However, due to the thin geometry of the standard accretion disk, the ionizing radiation from the thin disk may be modulated by $\cos \theta$, where $\theta$ is the viewing angle \citep[][]{Netzer1987}. This suggests that radiation anisotropy in sub-Eddington AGNs does exist, albeit not as pronounced as in super-Eddington AGNs. The geometry of the BLR is commonly described as a geometrically thick disk observed at a relatively face-on inclination \citep[e.g.,][]{Pancoast2014, Williams2018, Li2022}. Consequently, the ionizing radiation received by the observer and that received by the BLR clouds may differ by a factor of $\cos \theta_{\rm obs} / \cos \theta_{\rm BLR}$, where $\theta_{\rm obs}$ and $\theta_{\rm BLR}$ are the observer's line of sight and the viewing angles of the BLR with respect to the central engine, respectively. Assuming $\theta_{\rm obs} \sim 30^{\circ}$--$45^{\circ}$ and $\theta_{\rm BLR} \sim 45^{\circ}$--$75^{\circ}$, this factor ranges from approximately $1$ to $3.3$ (corresponding to $0$ to $0.5$ dex in logarithmic space). 

Observationally, the values of $\DeltaPhiH$ for AGNs at the lowest accretion rates (around $\log \Mdot \sim -2$, see Figure~\ref{fig:phiH_accretion_rate_delta}) are generally consistent with zero within the uncertainties. This indicates that $\PhiHiso \simeq \PhiHion$, suggesting that any anisotropy of the accretion-disk radiation is either not strong or effectively masked by measurement uncertainties.

\subsection{Anisotropic Radiation in super-Eddington AGNs}
\label{sec:Anisotropic Radiation}

Figure~\ref{fig:cartoon} presents a schematic representation of the BLRs illuminated by central engines in both sub- and super-Eddington AGNs. In the top panel, the ionizing photon flux encountered by the BLR is similar to or slightly smaller than that observed along the observer's line of sight in sub-Eddington AGNs ($\PhiHion \lesssim \PhiHiso$). This is because the value of $\Delta \log \PhiH$ lies between $0$ and $2$ for sub-Eddington AGNs (see Figure~\ref{fig:phiH_accretion_rate_delta}).

In contrast, in the super-Eddington regime (as shown in the lower panel of Figure~\ref{fig:cartoon}), the puffed-up inner region of the slim disk produces strong self-shadowing, reducing the ionizing continuum flux incident on the BLR and thus leading to $\PhiHion \ll \PhiHiso$. The reduced ionizing flux can also result in a smaller effective BLR size, as observed in super-Eddington AGNs \citep{du_supermassive_2015, du_supermassive_2016, Du_Supermassive_2018}, and a larger H$\beta$ responsivity. Specifically, the value of $\PhiHion$ is smaller than $\PhiHiso$ by factors of approximately $1.5$ to $3$ orders of magnitude in super-Eddington AGNs (see Figure~\ref{fig:phiH_accretion_rate_delta}). 

\begin{figure*}[!t]
    \centering
    \includegraphics[width=\textwidth]{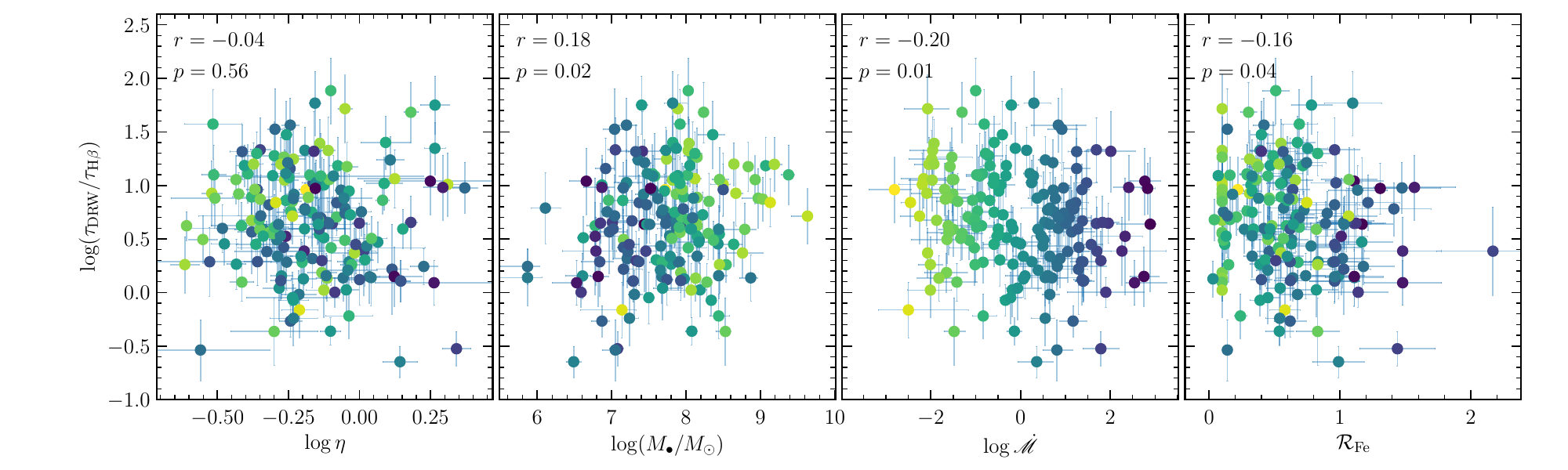} \\
    \includegraphics[width=\textwidth]{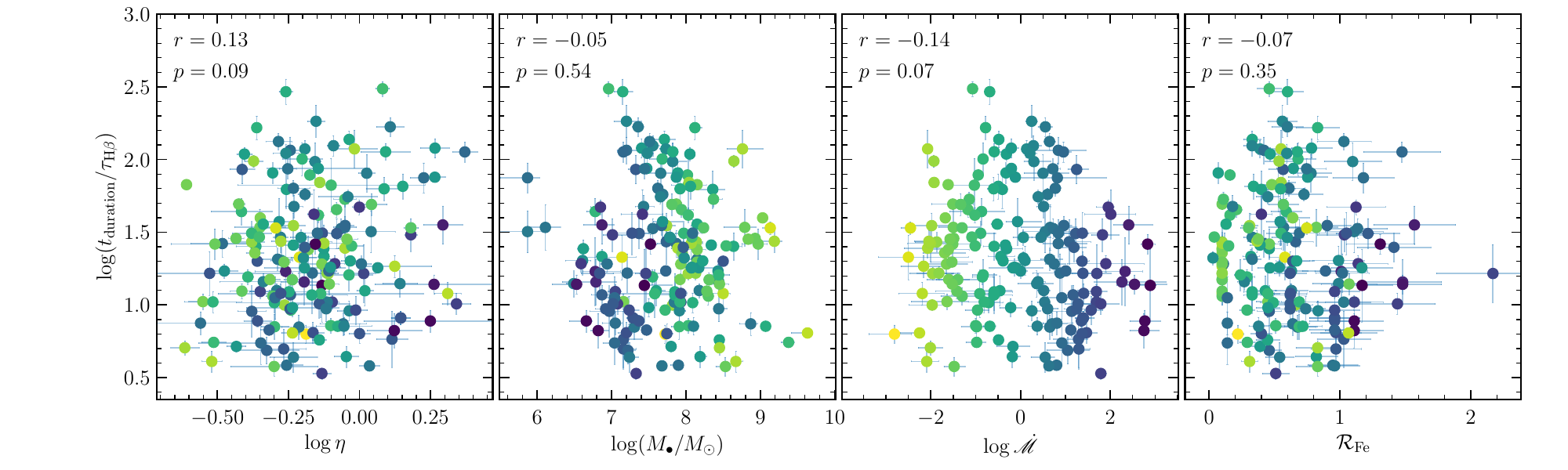}
    \caption{
    Correlations of $\log(\tau_{\mathrm{DRW}}/\tau_{\mathrm{H}\beta})$ (upper panels) and $\log(t_{\mathrm{duration}}/\tau_{\mathrm{H}\beta})$ (lower panels) with $\eta$, $M_{\bullet}$, $\Mdot$, and $\RFe$, color-coded by $\Mdot$. Here, $\tau_{\mathrm{DRW}}$ is the characteristic variability timescale of the continuum estimated from a damped random walk model, and $t_{\mathrm{duration}}$ is the duration of the monitoring campaign. The quantities $r$ and $p$ denote the Pearson correlation coefficient and its associated null probability, respectively.
    }
    \label{fig:logtau_DRW_duration_hb}
\end{figure*}

\subsection{Luminosity of H$\beta$}
\label{sec:luminosity}

If the radiation anisotropy in super-Eddington AGNs is sufficiently strong such that $\PhiHion$ is reduced relative to $\PhiHiso$ by 1.5 to 3 orders of magnitude, an important question naturally arises: is the ionizing radiation still powerful enough to photoionize the BLR clouds and produce H$\beta$ emission lines as strong as those observed in the spectra of super-Eddington AGNs?

Based on the \texttt{CLOUDY} photoionization calculations presented in Section~\ref{sec:PhiHion}, together with the observed range of $\eta$ for super-Eddington AGNs, we find that $\log (\PhiHion/{\rm cm^{-2}\,s^{-1}}) \approx 17.5$--$19.5$. In this regime, the equivalent width (EW) of the H$\beta$ line can reach $\sim 1000$--$1500$~\AA\ for directly illuminated BLR clouds (see also Figure~3g of \citealt{Korista1997} and Figure~1 of \citealt{Du2023_cloudy}). Given that $\PhiHion \sim 0.15$--$0.001\,\PhiHiso$, the expected observed H$\beta$ EWs in super-Eddington AGNs are therefore reduced to values of several to $\sim 200$~\AA. 

In practice, the measured H$\beta$ EWs in super-Eddington AGNs are typically in the range of $\sim 50$--$100$~\AA\ (see the compilation in \citealt{Du2019}), in good agreement with this simple estimate. This consistency indicates that, despite the strong anisotropy, the ionizing radiation remains sufficient to power the observed H$\beta$ emission.

Indeed, the H$\beta$ EWs of super-Eddington AGNs are observed to be systematically smaller than those of sub-Eddington AGNs \citep[e.g.,][]{Netzer2004, du_supermassive_2015, Du2023_cloudy}. This can be understood in the context of anisotropic ionizing radiation. The observed EW is measured relative to the continuum seen by the observer, which can be substantially stronger than that incident on the BLR. Consequently, the observed H$\beta$ EW is reduced relative to the value expected from the incident ionizing flux alone.

For the present sample, the average $\kappa$ of the super-Eddington AGNs is moderately larger than that of the sub-Eddington AGNs (6.6$\pm$0.8 versus 2.8$\pm$2.2), based on the accretion-disk model described in Section~\ref{sec:ionizing photons}. This implies that, for a given observed 5100~\AA\ continuum, super-Eddington AGNs are expected to have a stronger intrinsic ionizing continuum, making the substantial reduction of the ionizing flux incident on the BLR even more pronounced. Taken together, the systematically smaller observed H$\beta$ EWs and the larger $\kappa$ values provide further qualitative support for strong anisotropy of the ionizing radiation in super-Eddington AGNs.

\subsection{Systematic Effects on the Measured Responsivity}

\cite{Goad2014} showed that, when the BLR is spatially extended, the measured responsivity can be affected by both the characteristic variability timescale of the driving continuum and the duration of the monitoring campaign. To examine whether these effects could account for the larger responsivities observed in high-accretion-rate objects, we investigated the correlations of $\tau_{\rm DRW}/\tau_{\mathrm{H}\beta}$ with $\eta$, $M_{\bullet}$, $\dot{\mathscr{M}}$, and ${\cal R}_{\rm Fe}$ (Figure~\ref{fig:logtau_DRW_duration_hb}). Here, $\tau_{\rm DRW}$ is the characteristic variability timescale of the continuum estimated using a damped random walk model.

We find no significant correlation between $\tau_{\rm DRW}/\tau_{\mathrm{H}\beta}$ and any of $\eta$, $M_{\bullet}$, $\dot{\mathscr{M}}$, or ${\cal R}_{\rm Fe}$. We also examined the corresponding correlations using $t_{\rm duration}/\tau_{\mathrm{H}\beta}$, where $t_{\rm duration}$ is the duration of the monitoring campaign, and again find no significant correlation with any of these quantities (Figure~\ref{fig:logtau_DRW_duration_hb}). These results suggest that neither the characteristic variability timescale of the driving continuum nor the campaign duration is likely to be the primary cause of the systematically larger responsivities observed in high-accretion-rate AGNs.

The responsivity may also vary from season to season within an individual source \cite[e.g.,][]{Goad2004}. For sources with measurements from multiple monitoring seasons, we therefore adopt the responsivity averaged over the available seasons, which helps reduce the impact of seasonal fluctuations on the statistical trends considered here. Such variations may arise, at least in part, from changes in the characteristic variability timescale of the driving continuum. However, as shown above, we find no evidence that this variability-timescale effect is systematically related to either responsivity or accretion rate. Thus, although season-to-season variations may contribute additional scatter to the responsivity measurements of individual objects, they are unlikely to produce the systematic enhancement of responsivity observed in high-accretion-rate AGNs and therefore should not affect our main conclusion.

\subsection{Alternative Estimate of Responsivity}\label{sec:Fvar}

An alternative estimate of the responsivity can be obtained from the ratio of the fractional variability amplitudes of the emission line and the continuum, defined as
\begin{equation}
\eta_{F_{\rm var}}=\frac{F_{\rm var}({\rm H}\beta)}{F_{\rm var}({\rm cont})},
\end{equation}
\citep[e.g.,][]{Krolik1991, Goad2014}. Here, the fractional variability amplitude is defined as
\begin{equation}
F_{\rm var}=\frac{\left(S^{2}-\langle\sigma_{\rm err}^{2}\rangle\right)^{1/2}}{\langle F\rangle},
\end{equation}
where $S^{2}$ is the variance of the light curve, $\langle\sigma_{\rm err}^{2}\rangle$ is the mean squared measurement uncertainty, and $\langle F\rangle$ is the mean flux. We compare $\eta_{F_{\rm var}}$ with the effective responsivity obtained from the flux--flux relation after accounting for the continuum--emission-line time delay, as described in Section~\ref{sec:Responsivity}. The two estimates show a good correlation. Moreover, the correlations of the responsivity with $M_{\bullet}$, $\Mdot$, and $\RFe$ remain unchanged when $\eta_{F_{\rm var}}$ is adopted instead (see Appendix~D). This agreement indicates that our measurements of the effective H$\beta$ responsivity are not sensitive to the specific estimator adopted.

\subsection{Other Mechanisms}
\label{sec:other_mechanisms}

Motivated by the discovery of potential gravitational wave events in AGNs (GW~190521 and GW~231123; see \citealt{Graham2020, Abbott2020, Abac2025}), the existence of stellar-mass objects (e.g., very massive stars, compact objects, or stellar-mass black holes) in the accretion disks of AGNs as additional energy sources has recently gained attention \citep[e.g.,][]{Gilbaum2022, Zhou2024, Wang2025}. These extra energy sources may, on one hand, extend the continuum emitting region of the accretion disk and provide a possible alternative mechanism for the shortened H$\beta$ lags measured in super-Eddington AGNs \citep[see][]{Wang2025}. On the other hand, they could also increase the vertical scale of the accretion disk and enhance its anisotropy \citep{Gilbaum2022, Epstein-Martin2025}, particularly in the super-Eddington regime. This may contribute to the observed correlation of $\DeltaPhiH$ with accretion rate noted in the present paper, highlighting the need for detailed calculations in the future.

Other properties of the BLR may also affect the inferred $\PhiH$. For example, variations in the cloud column density or microturbulent velocity can alter the line emissivity and responsivity. These effects may therefore contribute to the observed trends and warrant further investigation with more detailed BLR photoionization models.

\subsection{Implications for Little Red Dots}
\label{sec:little_red_dots}

Recently, the JWST has discovered a peculiar class of objects at high redshifts ($4 \lesssim z \lesssim 10$) characterized by reddened rest-frame optical colors, compact morphologies, and ``V-shaped'' SEDs. These objects have been dubbed ``little red dots'' (LRDs) \citep[e.g.,][]{Labbe2023, Kocevski2023, Matthee2024, Greene2024}. They are primarily believed to host accreting SMBHs embedded in dense environments, as indicated by the detection of broad Balmer emission lines in their spectra. However, a range of detailed scenarios has been proposed to explain their observed properties \citep[e.g.,][]{Naidu2025, Wang2025LRD, Liu2025, Inayoshi2025, Zhang2026, Begelman2026, Chen2026}.

Applying local single-epoch mass estimates for sub-Eddington AGNs \citep{Greene2005}, the typical masses of LRDs are found to be $M_{\bullet} \sim 10^{6-8} M_{\odot}$ \citep{Inayoshi2025}. Nonetheless, it is suggested that the central SMBHs in LRDs accrete at super-Eddington rates, as indicated by their intrinsically X-ray weak SEDs \citep[e.g.,][]{Madau2024, Inayoshi2025b, Inayoshi2025, Lambrides2026}. Consequently, the radiation anisotropy associated with super-Eddington accretion may result in a reduction of their BLR sizes, implying that their SMBH masses could be overestimated by more than one order of magnitude \citep[e.g.,][]{Wang2014_self, du_supermassive_2015, Du2019, Yang2024, Lupi2024}. Furthermore, processes such as scattering by free electrons may broaden the intrinsic Balmer emission lines, further leading to an overestimation of their line widths and SMBH masses \citep[e.g.,][]{Rusakov2026}. Thus, more detailed considerations and more accurate mass estimates for the SMBHs in LRDs are essential for future research.

\begin{deluxetable}{cccc}[ht]
    \tablecaption{Partial correlation analysis of $\Delta\log\Phi_{\rm H}$ with $\M$ and $\Mdot$. Here $r^{\prime}$ denotes the partial correlation coefficient, and $p^{\prime}$ is the corresponding null probability. }\label{table:partial correlation}
\tablewidth{0pt}
\tablehead{ \colhead{Correlation} & {Confounding Variable} & {$r^{\prime}$} & {$p^{\prime}$}}
\centering
\startdata
\multicolumn{4}{c}{\rule{0pt}{3ex}  $\PhiHiso$($\kappa=2.2$), $\PhiHion$($Z=Z_{\odot}$)}\\
\hline
vs.$\log(\M/M_{\odot})$    &  $\log\Mdot$  &  -0.19  &  $3.50\times 10 ^{-2}$ \\
vs.$\log\Mdot$    &  $\log(\M/M_{\odot})$  &  0.38  &  $9.39\times 10 ^{-6}$ \\
\hline
\multicolumn{4}{c}{\rule{0pt}{3ex}  $\PhiHiso$($\kappa$ depends on $\M$ and $\Mdot$), $\PhiHion$($Z=5Z_{\odot}$)}\\
\hline
vs.$\log(\M/M_{\odot})$    &  $\log\Mdot$  &  -0.38  &  $1.47\times 10 ^{-5}$ \\
vs.$\log\Mdot$    &  $\log(\M/M_{\odot})$  &  0.57  &  $2.62\times 10 ^{-12}$ \\
\enddata
\end{deluxetable}

\section{Conclusion} \label{sec:Conclusion}

In this work, we systematically investigate the responsivity of broad H$\beta$ emission lines based on the light curves of RM objects, aiming to reveal how the strength of radiation anisotropy in the accretion disk depends on AGN properties. The main results are summarized as follows:

\begin{enumerate}[label=\arabic*.]

\item We found that the responsivity of H$\beta$ to continuum variability is statistically anti-correlated with SMBH mass and positively correlated with accretion rate and the strength of iron emission.

\item From \texttt{CLOUDY} photoionization calculations, we determined that the responsivity of BLR clouds is more sensitive to the incident ionizing photon flux $\PhiH$ than to the local gas density $\nH$ of the clouds. By comparing the ionizing photons ($\PhiHion$) received by the BLR clouds--derived from responsivity based on \texttt{CLOUDY} calculations--with those ($\PhiHiso$) inferred from the observer's line of sight under the isotropic assumption, we can investigate radiation anisotropy in different AGNs.

\item $\PhiHion$ is more similar to $\PhiHiso$ in sub-Eddington AGNs, while a greater discrepancy is observed in super-Eddington AGNs. The difference $\DeltaPhiH$ shows the most significant correlation with accretion rate, indicating a higher degree of radiation anisotropy in super-Eddington AGNs compared to sub-Eddington ones. This provides observational support for the self-shadowing effect expected from slim-disk accretion. 

\end{enumerate}

As discussed in Section \ref{sec:other_mechanisms}, compact objects or stellar-mass black holes within accretion disks, acting as additional energy sources, may further puff up the disks and enhance their radiation anisotropy, in addition to the \emph{self-shadowing} effect. Therefore, we will consider the contribution of these additional energy sources in the super-Eddington regime in future work.

\appendix
\section{Calculations of Ionizing Fluxes in Different Scenarios} \label{sec:appendix 5z}

\begin{figure*}[t]
    \centering
    \includegraphics[width=\textwidth]{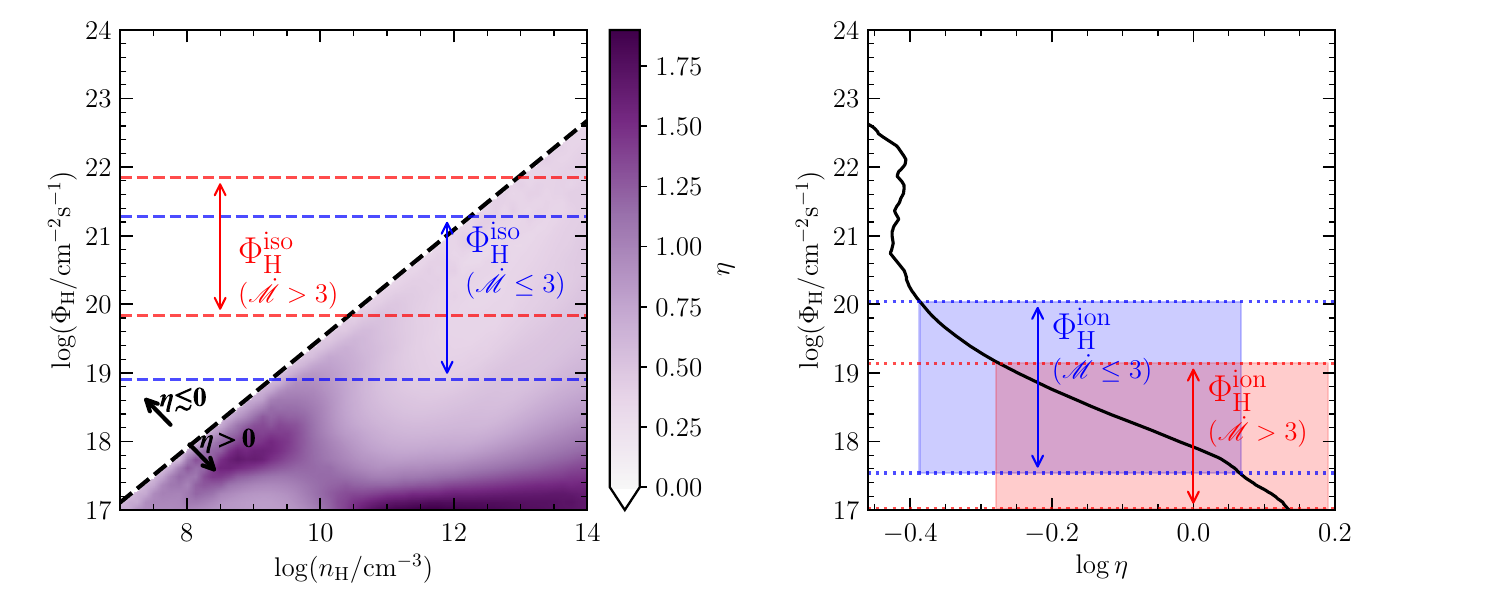}
    \caption{Same as Figure~\ref{fig:cloudy}, but for the scenario (under the single-cloud approximation) where $\kappa$ depends on $M_{\bullet}$ and $\Mdot$, utilizing \texttt{CLOUDY} calculations for a metallicity of $5Z_\odot$ and a super-Eddington SED \citep{ferland2020}. The meanings of the panels and lines are consistent with those in Figure~\ref{fig:cloudy}. }
    \label{fig:cloudy_5z}
\end{figure*}

\begin{figure*}[t]
    \centering
    \includegraphics[width=\textwidth]{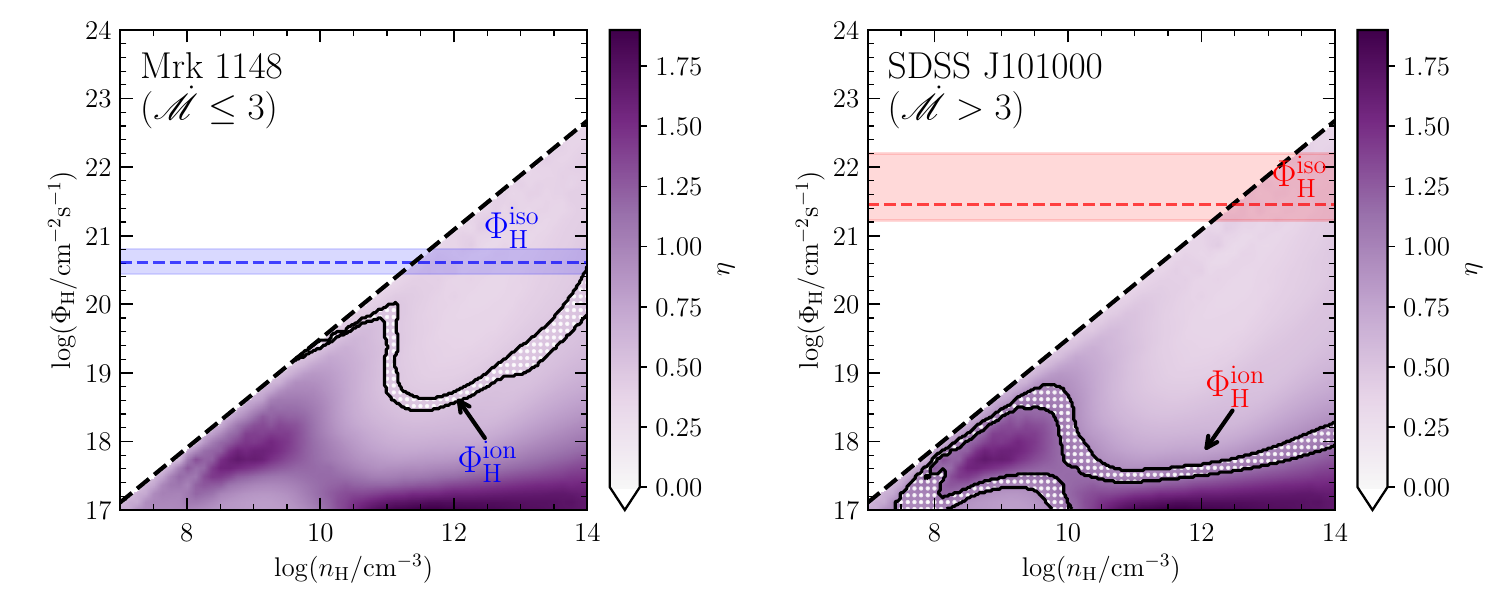}
    \caption{Same as Figure~\ref{fig:cloudy_examples}, but for the scenario (under the single-cloud approximation) where $\kappa$ depends on $M_{\bullet}$ and $\Mdot$, utilizing \texttt{CLOUDY} calculations for a metallicity of $5Z_\odot$ and a super-Eddington SED \citep{ferland2020}. The meanings of the panels and lines are consistent with those in Figure~\ref{fig:cloudy_examples}. }
    \label{fig:cloudy_examples_5z}
\end{figure*}

We calculated $\PhiHiso$ and $\PhiHion$ based on two scenarios: (1) a constant $\kappa$ along with \texttt{CLOUDY} calculations at solar metallicity and using the classic SED \citep{Mathews1987}, and (2) a $\kappa$ that depends on $\M$ and $\Mdot$, along with \texttt{CLOUDY} calculations for a metallicity of $5Z_\odot$ and a super-Eddington SED \citep{ferland2020}. However, in Figures \ref{fig:cloudy} and \ref{fig:cloudy_examples}, we only present the results from scenario (1). Here, we provide similar figures corresponding to scenario (2) in Figures \ref{fig:cloudy_5z} and \ref{fig:cloudy_examples_5z}.

\section{{Alternative LOC-based Estimate of $\PhiHion$ at the Responsivity-Weighted Radius}}
\label{sec:phihiso_RRW}

Following the definition of the responsivity-weighted radius $R_{\rm RW}$ in \citet{Bottorff2002} and using the relation $R\propto\Phi_{\rm H}^{-1/2}$, the corresponding ionizing photon flux can be written as
\begin{equation}
    \Phi_{\rm H,RW}(\gamma, \delta)=
    \left[\frac{\displaystyle\iint \Phi_{\rm H}^{-1/2}W_{\rm RW}\,
     d \log n_{\rm H}\, d \log\Phi_{\rm H}}
    {\displaystyle\iint W_{\rm RW}\, d \log n_{\rm H}\, d \log\Phi_{\rm H}}\right]^{-2},
    \label{eq:phi_LOC_RW}
\end{equation}
where $W_{\rm RW}=\eta(\Phi_{\rm H},n_{\rm H}) F_{\mathrm{H}\beta} (\Phi_{\rm H},n_{\rm H})
\Phi_{\rm H}^\gamma n_{\rm H}^{\delta+1}$. 

\begin{figure*}
    \centering
    \includegraphics[width=\textwidth]{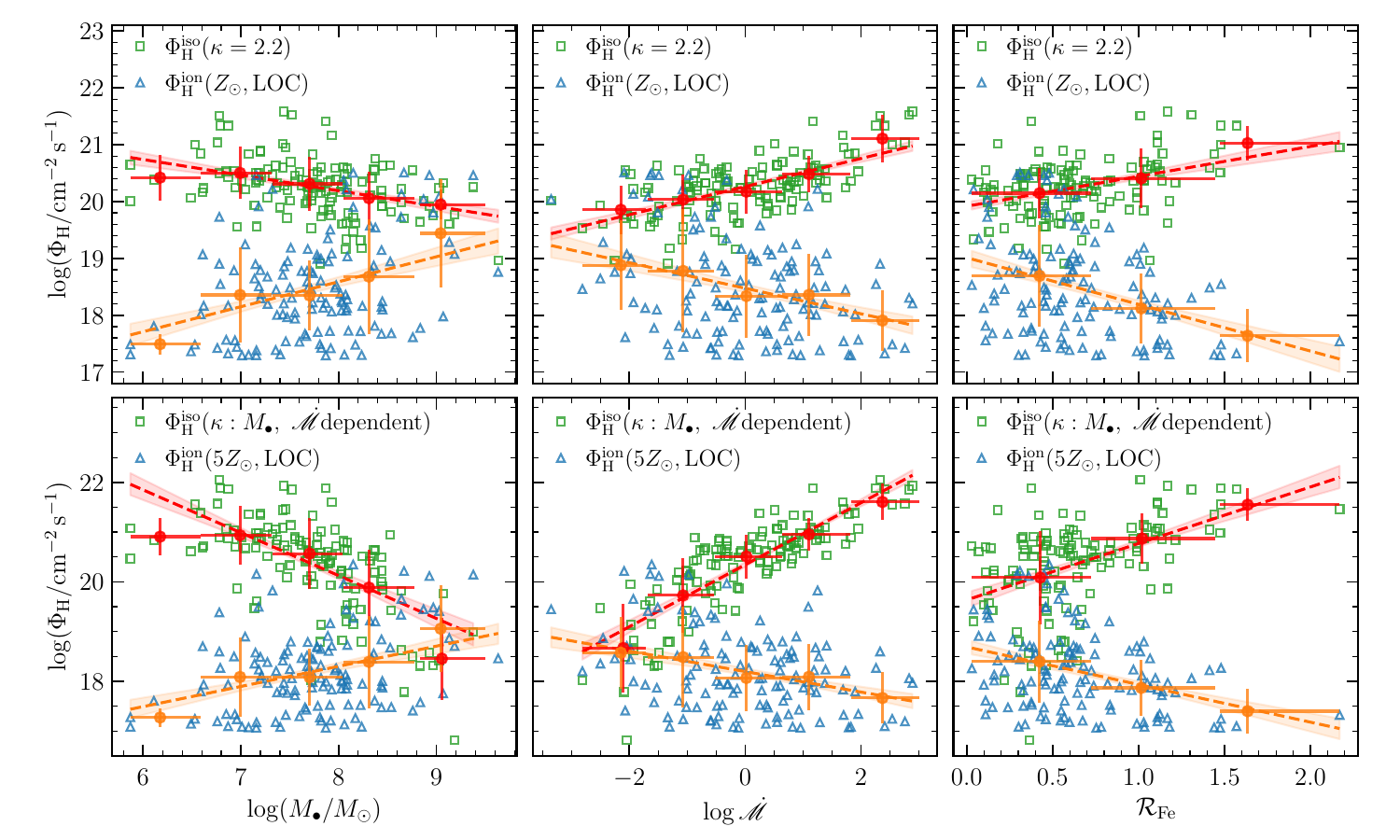}
    \caption{
    Same as Figure~\ref{fig:phiH_accretion_rate_LOC_EW}, but for the LOC-based $\PhiHion$ inferred from the responsivity-weighted radius. The legend and symbols have the same meanings as in Figure~\ref{fig:phiH_accretion_rate_LOC_EW}.} 
    \label{fig:phiH_accretion_rate_LOC_RRW}
\end{figure*}

\begin{figure*}[!t]
    \centering
    \includegraphics[width=\textwidth]{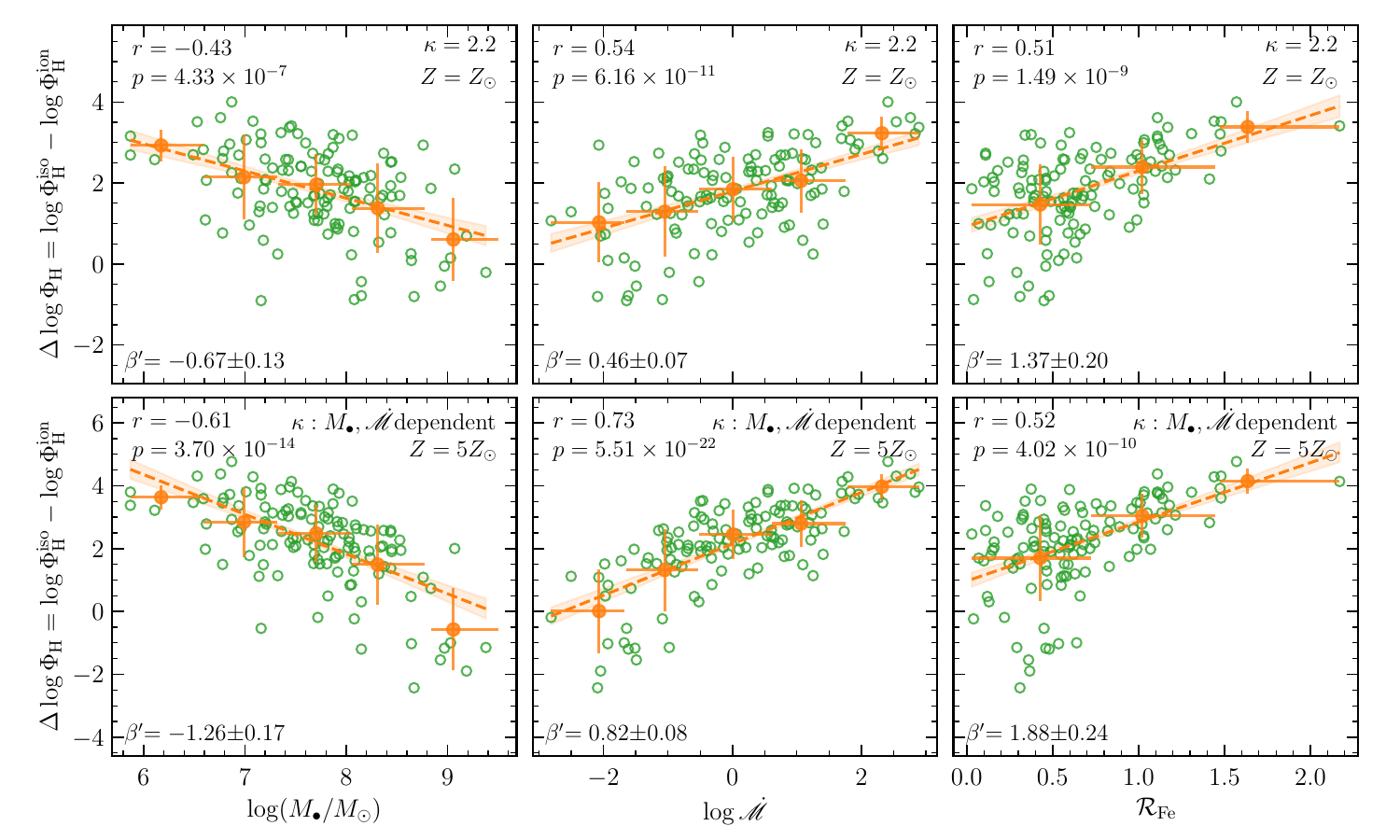}
    \caption{
    Same as Figure~\ref{fig:phiH_accretion_rate_delta}, but for the LOC-based $\DeltaPhiH$ inferred from the responsivity-weighted radius. The upper and lower panels correspond to the two scenarios as shown in Figure~\ref{fig:phiH_accretion_rate_LOC_RRW}. The legend and symbols have the same meanings as in Figure~\ref{fig:phiH_accretion_rate_delta}.}
    \label{fig:LOC_phi_H_all_delta_RRW}
\end{figure*}

We adopt the same parameter ranges, $-2.5\leq\gamma\leq0$ and $-1.5\leq\delta\leq-0.5$, and the same allowed $\gamma$--$\delta$ region for each source as in Section~\ref{sec:phihion_eta_LOC}. The median value of $\Phi_{\rm H,RW}(\gamma,\delta)$ over the corresponding allowed region is adopted for each source. The resulting $\PhiHion$ values are presented in Figure~\ref{fig:phiH_accretion_rate_LOC_RRW}, while the corresponding $\Delta\log\Phi_{\rm H}$ are shown in Figure~\ref{fig:LOC_phi_H_all_delta_RRW}. Adopting the flux at $R_{\rm RW}$ preserves the correlations of $\PhiHion$ and $\Delta\log\Phi_{\rm H}$ with $M_{\bullet}$, $\dot{\mathcal{M}}$, and $R_{\rm Fe}$. 

\section{LOC-weighted Responsivity}
\label{sec:appendix LOC-weighted responsivity}

Here we recalculated the $\eta(\PhiH)$ relation (the right panel of Figure~\ref{fig:cloudy}) using the LOC model. Following the definitions and formalism of \citet{Bottorff2002}, the effective responsivity at a given $\PhiH$ is weighted by the H$\beta$ surface flux and the adopted LOC covering-factor distribution, i.e.,
\begin{equation}
\eta(\PhiH)=\frac{\displaystyle\int \eta(\PhiH,\nH) F_{\rm H\beta}(\PhiH,\nH) \frac{\partial^2C_{\rm f}}{\partial \nH \partial \PhiH}\,d\nH}{\displaystyle\int F_{\rm H\beta}(\PhiH,\nH) \frac{\partial^2C_{\rm f}}{\partial \nH \partial \PhiH}\,d\nH},
\end{equation}
where $\partial^2C_{\rm f}/\partial \nH \partial \PhiH$ is assumed to follow $\partial^2C_{\rm f}/\partial \nH \partial \PhiH \propto f(\PhiH)g(\nH) \propto \PhiH^{\gamma}\nH^{\delta}$. For simplicity and for demonstration, here we adopt $g(\nH)\propto \nH^{-1}$. Therefore the corresponding LOC-weighted responsivity is
\begin{equation}
\eta(\PhiH)=\frac{\displaystyle\int\eta(\nH,\PhiH)F_{\rm H\beta}(\nH,\PhiH)\,d\log\nH}{\displaystyle\int F_{\rm H\beta}(\nH,\PhiH)\,d\log\nH}.
\end{equation}
This weighting reflects the fact that the effective line responsivity is determined by the differential response of the integrated line emission and is therefore weighted by the H$\beta$ emission contribution from clouds with different physical conditions.

\begin{figure}[h]
    \centering
    \includegraphics[width=0.9\textwidth]{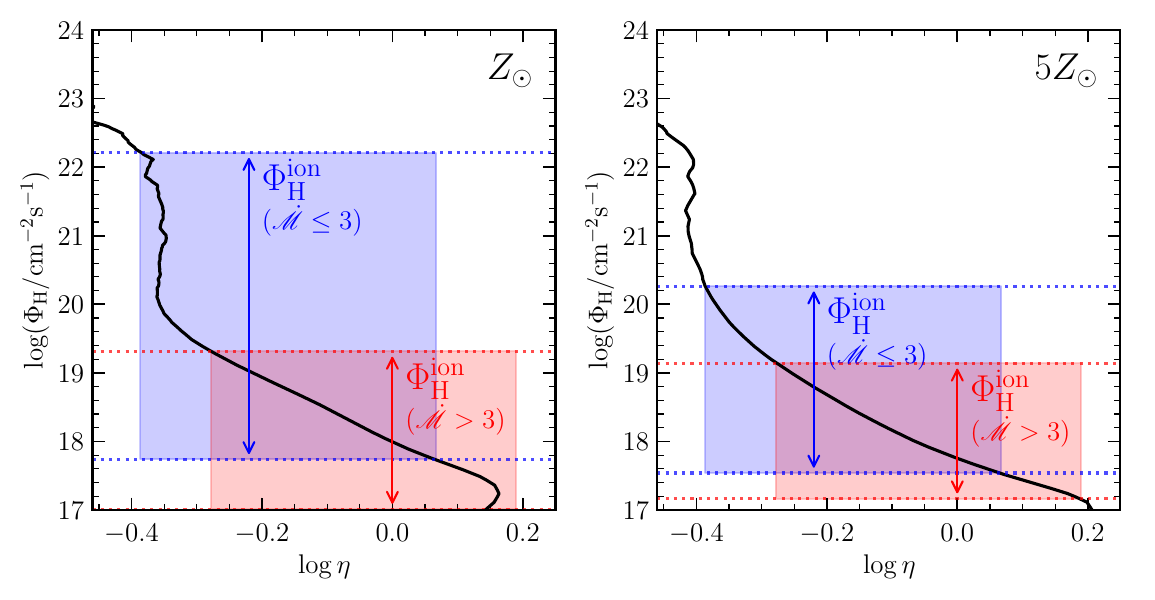}
    \caption{Same as the right panel of Figure~\ref{fig:cloudy}, but with the black lines showing the LOC-weighted $\eta(\PhiH)$ obtained by adopting $g(\nH)\propto\nH^{-1}$. The meanings of the panels and lines are the same as those in Figure~\ref{fig:cloudy}. }
    \label{fig:cloudy_flux_weighted_Z_5Z}
\end{figure}

We perform this test for both the low- and high-metallicity \texttt{CLOUDY} grids. As shown in Figure~\ref{fig:cloudy_flux_weighted_Z_5Z}, the LOC-weighted $\eta(\PhiH)$ relations are broadly consistent with those obtained under the simple-average assumption (Section~\ref{sec: PhiHiso and PhiHion}). In particular, the H$\beta$ responsivity still increases toward lower incident ionizing photon flux. More importantly, the comparison between the inferred $\PhiHion$ and the corresponding $\PhiHiso$ remains essentially unchanged: the $\PhiHion$ regions remain clearly offset from the $\PhiHiso$ regions. These results indicate that, although the quantitative form of the inferred $\eta(\PhiH)$ relation can depend on the adopted weighting scheme, our main conclusions are robust against the treatment of the cloud distribution.

\section{Comparison of Different Responsivity Measurements}

\begin{figure}[h]
    \centering
    \includegraphics[width=0.38\textwidth]{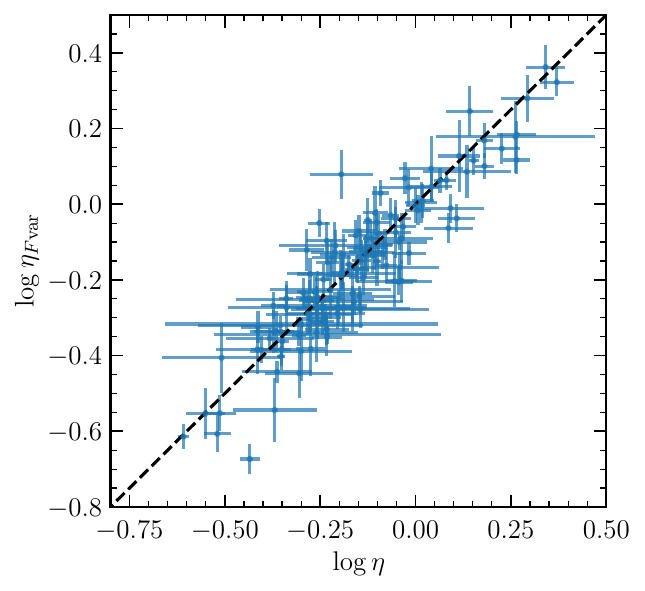}
    \caption{Comparison between the responsivity used in the main text and that estimated from the fractional-variability ratio, $\eta_{F_{\rm var}}=F_{\rm var}({\rm H}\beta)/F_{\rm var}({\rm cont})$. The dashed line denotes equality between the two estimates. Multiple measurements of the same object were combined. }
    \label{fig:logeta_comparison}
\end{figure}

\begin{figure}[h]
    \centering
    \includegraphics[width=0.9\textwidth]{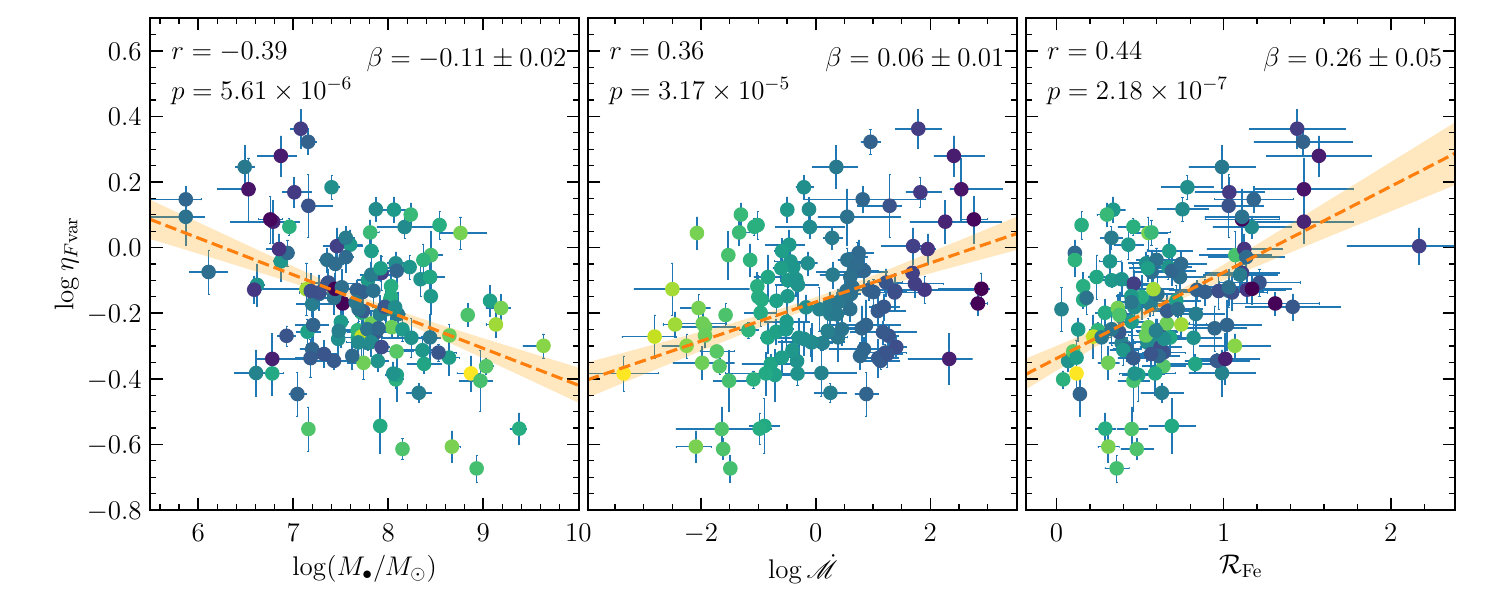}
    \caption{Same as Figure~\ref{fig:logeta_M_RFe}, but using the responsivity estimated from the ratio of the fractional variability amplitudes, $\eta_{F_{\rm var}}=F_{\rm var}({\rm H}\beta)/F_{\rm var}({\rm cont})$. The legend and symbols have the same meanings as in Figure~\ref{fig:logeta_M_RFe}.}
    \label{fig:logeta_Fvar}
\end{figure}

An alternative estimate of the responsivity can be obtained from the ratio of the fractional variability amplitudes of the emission line and the continuum ($\eta_{F_{\rm var}}$, see Section~\ref{sec:Fvar}). As shown in Figure~\ref{fig:logeta_comparison}, $\eta_{F_{\rm var}}$ is broadly consistent with the effective responsivity obtained using the flux--flux relation described in Section~\ref{sec:Responsivity}. We further repeat the analysis presented in Figure~\ref{fig:logeta_M_RFe}, replacing the regression-based responsivity with $\eta_{F_{\rm var}}$. As shown in Figure~\ref{fig:logeta_Fvar}, the correlations of the responsivity with $M_{\bullet}$, $\Mdot$, and $\RFe$ remain almost unchanged. These results demonstrate that our main conclusions are not sensitive to the particular estimator adopted for the effective responsivity.

\section*{ACKNOWLEDGEMENTS}
We thank the anonymous referee for constructive comments and suggestions that helped improve the manuscript. This research is supported by the National Science Foundation of China through grants NSFC-12521005, NSFC-12641301, NSFC-12333003, the National Key R\&D Program of China (2023YFA1607903 and 2021YFA1600404), and the National Science and Technology Major Project (2024ZD0300303). Additionally, it is funded by the China Manned Space Project (CMS-CSST-2025-A07).

\software{BCES \citep{Akritas1996_BCES, Nemmen2012_BCES}, DASpec \citep{Du2024}. }

\FloatBarrier

\end{document}